\documentclass[fleqn,usenatbib]{mnras}
\usepackage{blindtext}
\usepackage{physics}
\usepackage[english]{babel}
\usepackage[utf8x]{inputenc}
\usepackage[T1]{fontenc}
\usepackage{txfonts}
\usepackage{aecompl}

\usepackage{amsmath}
\usepackage{graphicx}
\usepackage[colorinlistoftodos]{todonotes}
\usepackage{subfig}
\begin{document}

\label{firstpage}

\title{Time dependent spectral modeling of Markarian 421 during a violent outburst in 2010}
\author[B. Banerjee et al.]{
B. Banerjee$^{1}$\thanks{E-mail: biswajit.banerjee@saha.ac.in}, 
M. Joshi$^{2}$\thanks{E-mail: mjoshi@bu.edu},
P. Majumdar$^{1}$,
K. E. Williamson$^{2}$, 
S. G. Jorstad$^{2,3}$,
A. P. Marscher$^{2}$
\\
$^{1}$Saha Institute of Nuclear Physics, HBNI, 1/AF Bidhannagar, Salt Lake, Sector-1, Kolkata 700064, India\\
$^{2}$Institute for Astrophysical Research, Boston University, 725 Commonwealth Avenue, Boston, MA 02215\\
$^{3}$Astronomical Institute, St. Petersburg University, Universitetskij Pr. 28, Petrodvorets, 
198504 St. Petersburg, Russia}

\date{Accepted 2019 April 30; Received 2019 April 26; in original form 2019 January 21}
\pagerange{\pageref{firstpage}--\pageref{lastpage}}
\maketitle

\maketitle

\begin{abstract}
We present the results of extensive modeling of the spectral energy distributions (SEDs) of the closest blazar (z=0.031) Markarian 421 (Mrk 421) during a giant outburst in February 2010. The source underwent rapid flux variations in both X-rays and very high energy (VHE) gamma-rays as it evolved from a low-flux state on 2010 February 13-15 to a high-flux state on 2010 February 17. During this period, the source exhibited significant spectral hardening from X-rays to VHE gamma-rays while exhibiting a "harder when brighter" behavior in these energy bands. We reproduce the broadband SED using a time-dependent multi-zone leptonic jet model with radiation feedback. We find that an injection of the leptonic particle population with a single power-law energy distribution at shock fronts followed by energy losses in an inhomogeneous emission region is suitable for explaining the evolution of Mrk 421 from low- to high-flux state in February 2010. The spectral states are successfully reproduced by a combination of a few key physical parameters, such as the maximum $\&$ minimum cutoffs and power-law slope of the electron injection energies, magnetic field strength, and bulk Lorentz factor of the emission region. The simulated light curves and spectral evolution of Mrk 421 during this period imply an almost linear correlation between X-ray flux at 1-10 keV energies and VHE gamma-ray flux above 200 GeV, as has been previously exhibited by this source. Through this study, a general trend that has emerged for the role of physical parameters is that, as the flare evolves from a low- to a high-flux state, higher bulk kinetic energy is injected into the system with a harder particle population and a lower magnetic field strength.
\end{abstract}


\section{Introduction}

Blazars are considered as 
among the most 
violently variable
objects in the entire electromagnetic spectrum 
in the universe, 
explained by emission beamed along the line of sight by relativistic jets of high-energy plasma (\citealt{1979ApJ...232...34B}). Variability, on time scales of months to hours (and in some cases down to minutes), is a common phenomenon for these objects,
and is pervasive
throughout the electromagnetic spectrum. The spectral energy distribution (SED) of a blazar shows a double hump structure. Based on the position of the peak of the 
first hump, these
objects are characterized into four different classes: LBL (low-frequency peaked BL 
Lacertae
(BL Lac) object) peaking in the infrared to optical regime, IBL (intermediate-frequency peaked BL Lac objects) peaking at optical- near-UV frequencies, HBL (high-frequency peaked BL Lac objects) peaking at X-ray energies, and FSRQs (flat-spectrum radio quasars) peaking in the infrared regime and exhibiting strong emission lines in their optical spectrum (\citealt{PG}; \citealt{Sambruna}; \citealt{Marscher}; \citealt{Konigl}; \citealt{Ghisellini}; \citealt{B2012}).

The origin of the double hump structure of the SED of blazars can be explained by leptonic and/or hadronic models. According to these models, the first hump is a result of synchrotron radiation of primary electrons accelerated to relativistic energies in the emission region. On the other hand, the production of the second hump 
changes, depending
on the model that is being used to explain its origin. In the case of a hadronic model, if relativistic protons are present in the jet of a blazar and have energies above the interaction threshold, then the second hump is a result of interactions between accelerated protons \& 
electron-positron pair cascades
and synchrotron photons responsible for the low-energy hump (\citealt{SPB2001}; \citealt{SPB2003}; \citealt{2010arXiv1006.5048B}). 
However, under
a leptonic scenario, the high-energy hump is a result of inverse Compton (IC) scattering 
of an internal
and/or external photon field by primary electrons. If the internal photon field, which is the synchrotron radiation in the emission region, is involved in IC 
scattering, then
the resultant emission is known as synchrotron self-Compton (SSC) 
(e.g.,
\citealt{1996Alan}). On the other hand, if an external photon field is 
involved, then
the corresponding emission is called external Compton (EC). The sources of this external photon field could be the photons from the accretion disk (\citealt{1992Dermer}; \citealt{1993Dermer}), or the broad line region (BLR) and/or the dusty torus (\citealt{SSC1}; \citealt{2014ApJ...785..132J}; \citealt{Sikora1994}). %

Mrk 421 is among the best studied 
BL Lac
objects 
at all wavelengths.
Rapid flux variations ($\sim$ minutes to hours) 
have been detected from the source by 
a number of instruments
(see \citealt{Gaidos}). 
The broad-band SED of the source has been measured in great detail by most of the current generation instruments
at a number of epochs.
At TeV energies, the source has displayed many high-flux states (\citealt{14Y}). It was detected in a flaring state by 
the
Whipple telescope in 1995 (\citealt{1995Malcomb}). 
These authors
reported the first ever quasi-simultaneous multi-waveband observations of this source. The event was characterized by a quiescent or a steady flux at optical and 100 MeV-GeV regimes while the source varied strongly at X-ray and TeV energies. In 1997, another flaring event was observed, where the highest flux state was detected by Whipple telescope to be around ten times that of the 
Crab Nebula
above 500 GeV (\citealt{Whipple19971}; \citealt{Whipple2}). An outburst with a flux level of 13 
Crab units
(above 380 GeV) and high variability of the source was reported in 2001 by Whipple (\citealt{2001W}), where the flux state remained high from January to May 2001. Another big flaring event was reported by Whipple ($>$ 400 GeV) (\citealt{2008ATel}) and VERITAS ($>$ 300 GeV) (\citealt{VERITAS200608}) 
in a 2003-2004 campaign,
where the flux reached 10 Crab units. Similarly, in X-rays the source has displayed many outbursts. An orphan flaring event (X-ray flares with no VHE counterpart) was reported in 2003-2004 campaign by \citealt{Fraija}. The source showed another violent outburst in 2006 in X-rays, where the measured flux reached $\sim$ 
85 milli-Crab
(mCrab) in the energy range of 2-10 keV (\citealt{Tramacere}; \citealt{Ushio}). The peak of the synchrotron hump appeared beyond 10 keV for this flaring event. 
Super-AGILE,
operational in the hard 
X-rays, reported
a flare on 2008 June 10 (\citealt{2008ATelAGILE}).
However, the strongest hard X-ray flare reported by MAXI (Monitor of ALL-sky X-ray Image) was in February 2010, where the flux went up to $\sim$ 164 $\pm$ 17 mCrab (\citealt{I2010}; \citealt{I2015}). This is the largest X-ray flare ever reported for this source (\citealt{2012Niinuma}). In addition, it was followed by 
very bright
VHE $\gamma$-ray emission ($\sim$ 10 Crab) (Ong, R. A. 2010, ATel, 2443, 1).\\
In this paper, we focus on analyzing the spectral states of Mrk 421 that were observed in February 
2010, during which the blazar was found to be in one of the brightest states in both X-ray and TeV energies (\citealt{VERITAS}; \citealt{HAGAR}; \citealt{TACTIC}; \citealt{ARGO}; \citealt{HESS}). Such flaring events offer a unique possibility of understanding the role of particle acceleration along with the temporal evolution of the source from a low- to a high-flux state. A similar but low magnitude ($\sim$ 2 Crab flux above 200 GeV) flare was detected with unprecedented multi-wavelength (MWL) data coverage during March 2010 (\citealt{March2010}). Through this study, we aim to constrain the role of intrinsic parameters and understand the physical processes in Mrk 421 responsible for producing different flux states during the giant outburst of February 2010. We reproduce a total of three spectral states observed over a period of five consecutive days, from 13-17 February 2010, using a multi-zone time-dependent leptonic jet model with radiation feedback (\citealt{Joshi2011}). 
These days include a
representative (a) low-flux state (from 2010 February 13-15), (b) intermediate-flux state (2010 February 16), and (c) high-flux state (2010 February 17). In order to establish a baseline for the nature of relevant physical parameters for the source and facilitate the comparison of parameters between the three spectral states, we also reproduce 
an averaged emission state
of Mrk 421. This emission, which we term as the steady state (SS), is reported in \citealt{SteadyState} and has been obtained by averaging the flux of Mrk 421 over a period of 4.5 months when the source was found in a relatively quiescent state.

We have used a cosmology with $\Omega_m$  = 0.3, $\Omega_{\Lambda}$ = 0.7, and $H_0$ = 70 $km$ $s^{-1}$ $Mpc^{-1}$. In this cosmology, for a redshift of $z$ = 0.031, the luminosity distance of Mrk 421 is $dL$ = 136 $Mpc$.

The paper is organized as follows: 
we discuss our data
collection and source activity during this period in \S 2. We describe the model in \S 3. The key findings are discussed in \S 4 . 
Finally, we
summarize the results and 
discuss future possibilities for probing the physics of Mrk 421
in \S 5.

\section{Data collection:}

Several collaborations 
like VERITAS
(\citealt{VERITAS}), TACTIC (\citealt{TACTIC}; \citealt{TACTIC2}), HAGAR (\citealt{HAGAR}, ARGO-YBJ (\citealt{ARGO}), and H.E.S.S. (\citealt{HESS})
reported that Mrk 421 underwent a huge flare in TeV and X-ray energies in February 2010, exhibiting rapid changes in flux and spectral index from 2010 February 13 (MJD 55240) to 17 (MJD 55244). 
Figure \ref{fig:MWL} shows the SEDs for the three spectral states 
under
consideration, along
with the SS data of Mrk 421 (\citealt{SteadyState}).
As discussed below, the low-flux state uses X-ray data from \citealt{HAGAR}, which is 
3-day
averaged data from 2010 February 13-15. Since Mrk 421 did not show any significant variation in X-rays over this time 
period, we
consider this averaged emission to be representative of 
a low-flux state over a 24-hour period.
As can be seen from Figure \ref{fig:MWL}, the low-flux state is different from SS in its flux and spectral hardness for both X-ray and $\gamma$-ray bands. This implies that the low-flux state represents the beginning of the flare in Mrk 421 that culminated in a high-flux state on 2010 February 17 in those bands. In addition, the source exhibited significant spectral variability in both X-ray and TeV regimes during intermediate- and high-flux states. However, the same level of variability was not observed at optical and GeV energies during that time. 
 The source was reported to be found in the highest flux and hardest spectral state on 2010 February 17 by both VERITAS and HAGAR 
in the 
VHE band (\citealt{VERITAS}; \citealt{HAGAR}), and the flaring
activity was also prominent in soft and hard X-rays (\citealt{HAGAR}). In this section, we summarize our multi-wavelength data collection.

\subsection{Optical and Ultraviolet data: }

We have used the optical data for Mrk 421 from Steward observatory. The data was collected in V band using 2.3 m Bok telescope on Kitt Peak, Arizona. 
The data was reduced
based on the procedure described in \citealt{Smith}. In addition, we have collected the optical and ultraviolet (UV) data for Mrk 421 from the UVOT telescope (\citealt{Roming}) on board \textit{Swift} observatory. Details on data reduction procedure are provided in \citealt{Williamson}. The UVOT data used for 2010 February 15 and 16 are from V-band, \textit{Swift}-UVW1, -UVW2, and -UVM2 filters. Data used for SED construction on 2010 February 17 are from V-band, UVW2, and UVM2 filters. As described in \citealt{Williamson}, all our data has been corrected for dereddening and host galaxy contribution. The dereddened  magnitudes for Swift and ground-based observations have been converted to fluxes based on the details given in the paper. 

\subsection{X-ray Data:}

MAXI (operational in 2 to 20 KeV), \textit{\textit{Swift}}-XRT (operational in 0.3 to 10 KeV) and \textit{\textit{Swift}}-BAT (operational in 20 to 60 KeV) observed this giant flaring activity in the soft and hard X-rays (\citealt{HAGAR}; \citealt{TACTIC}). Both papers reported that the flux measured by the four instruments (\textit{RXTE}-ASM [1.5-12 keV], -PCA [2-20 keV]; \& \textit{Swift}-XRT [0.5-2 KeV], -BAT [15-50 keV]) showed an increase by at least a factor of 4 within a period of two days as the source evolved from a low-flux to a high-flux state.
The activity in the hard X-ray band can be seen in 
light curves (LCs) 
presented in \citealt{HAGAR} from MAXI and \textit{\textit{Swift}}-BAT, which makes this event exceptional. The corresponding evolution of the synchrotron peak position in the observed SEDs can be seen in the inset of Figure \ref{fig:MWL}. As the peak position of the synchrotron component evolves 
from the low to the
high-flux state, the slope of the line joining the optical to X-ray data 
changes, indicating
spectral hardening in the X-ray regime during that time period. The inset of Figure \ref{fig:MWL} shows the change in slope of the line joining the optical and X-ray data for different flux states.\\
\subsection{High Energy $\gamma$-ray data from \textit{Fermi}-LAT:}
\textit{Fermi}-LAT (\citealt{Atwood}) is a pair conversion telescope, with a field of view (FoV) of above 2 sr, operating in the energy range from 20 MeV to 300 GeV. It is 
the most sensitive instrument
available in this energy range (\citealt{Ackermann}). 
A few months
after the launch in June 2008, \textit{Fermi}-LAT started to operate in all sky survey mode. The telescope 
scans
the whole sky in 3 hours (\citealt{Atwood}). 
For this paper
we have analyzed the Pass8\footnote{https://fermi.gsfc.nasa.gov/ssc/data/analysis/documentation\\/Pass8\_usage.html} data from 2010 February 13 to 17. We analyzed the data for Mrk 421 using the latest Fermi Science Tool\footnote{ https://fermi.gsfc.nasa.gov/ssc/data/analysis/software/} software package version v10r0p5. In order to determine the flux and the spectrum of the source, maximum likelihood optimization has been used (\citealt{Abdo200918346}). The data selection and quality checks have been made using the gtselect tool\footnote{https://fermi.gsfc.nasa.gov/ssc/data/analysis/scitools\\/likelihood\_tutorial.html}. Since the telescope is sensitive to $\gamma$-rays from the interactions of cosmic rays with the ambient matter, we set our maximum zenith angle  at 105 degrees to 
remove
the background $\gamma$-ray events from Earth's limb. The analysis 
included
all photons from a circular region of 10 degrees around 
Mrk 421, which
we call the region of interest (ROI). 
Only photons of energy above 100 MeV were considered for further analysis. The latest LAT instrument response function (IRF) ``P8R2\_SOURCE\_V6'' has been used. The third \textit{Fermi}-LAT catalog (3FGL catalog: \citealt{Acero}) has been used to include the contributions of sources inside the ROI. The spectral model of the source has been considered as a simple power law of the form 
$dN(E)/dE = N_0 (E/E_0)^{-\Gamma_{ph}}$, where $N_0$ is called the prefactor, 
$\Gamma_{ph}$ is the index,
and $E_{0}$ is the scale in energy.
The spectral parameters of the sources including Mrk 421 inside the ROI are kept 
free, whereas
the spectral parameters of the sources beyond 10 
degrees from Mrk 421
are kept fixed to the values according to the 3FGL catalog. The unbinned likelihood\footnote{https://fermi.gsfc.nasa.gov/ssc/data/analysis/scitools\\/likelihood\_tutorial.html} method has been used in order to estimate the detection significance of the sources. The test statistics parameter determined from the aforementioned method is given by $TS = 2 \Delta log(L)$, where L denotes the likelihood function between the model with the source and without the source. According to the definition TS=9 corresponds to a detection significance of $\sim 3 \sigma$ (\citealt{Mattox}). All the sources with TS < 9 are excluded from the likelihood analysis. In order to model the spectrum of sources we used the latest Galactic diffuse emission model ``gll\_iem\_v06'' and the isotropic background model ``iso\_P8R2\_SOURCE\_V6\_v06''. We estimated the corresponding butterfly plots for the source using the methods mentioned in the \textit{Fermi}-LAT webpage\footnote{https://fermi.gsfc.nasa.gov/ssc/data/analysis/scitools\\/python\_tutorial.html}. VERITAS observed the high-flux state of Mrk 421 for $\sim$ 6 hours. In order to obtain 
quasi-simultaneous
data with VERITAS, we also analyzed the Fermi-LAT data for this state by taking 12 hours around the VERITAS observation period. In Figure \ref{fig:SEDs}(d), we show the butterfly plots for both 24 \& 12 hours by closed butterfly (gray) and open butterfly with cross at the edges (blue), respectively, for the high-flux state.
\subsection{Very High Energy (VHE) gamma-ray data:}

\begin{figure*}
\includegraphics[width=0.7\textwidth,angle=270]{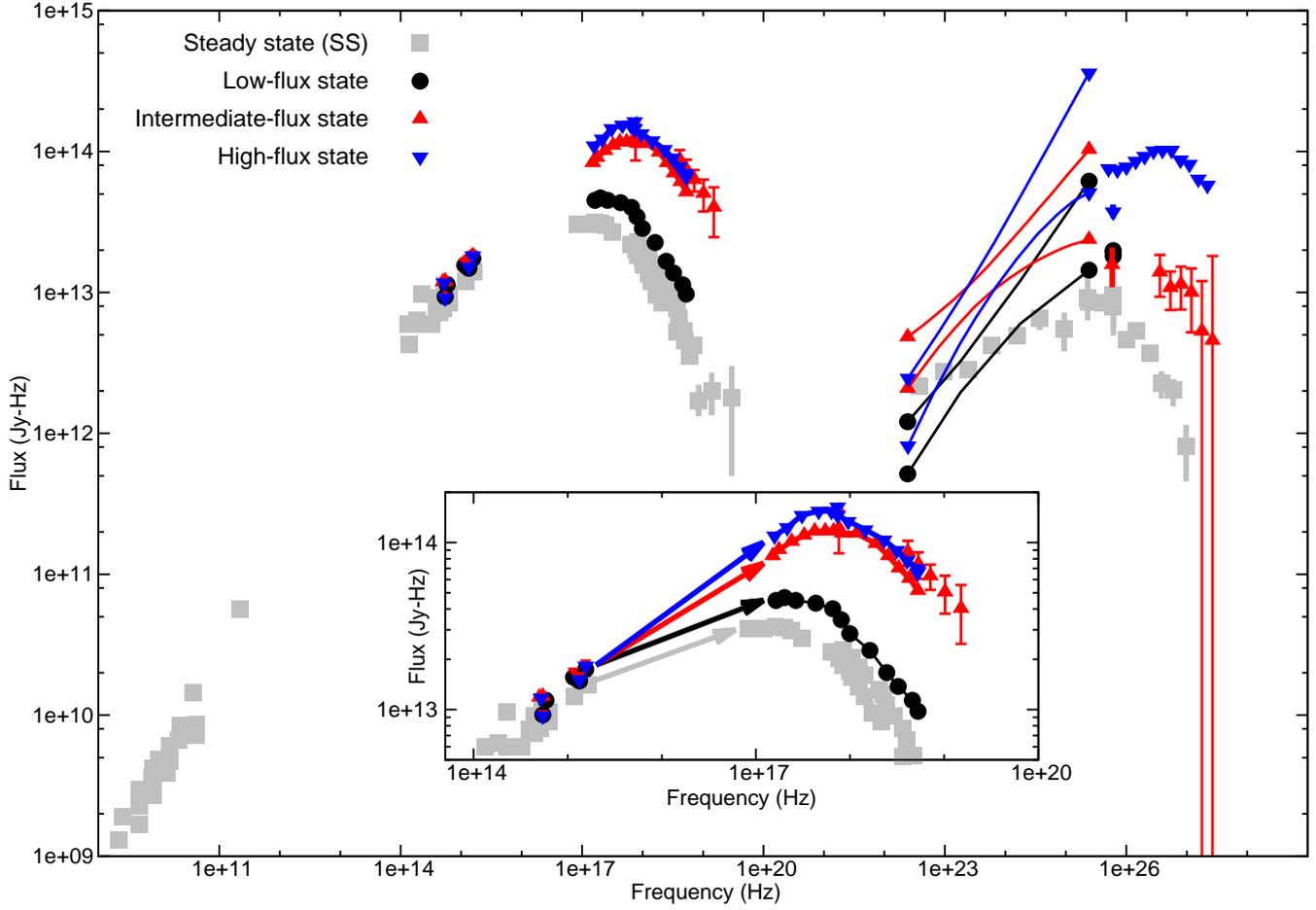}
\caption{
Multi-wavelength
data of Mrk 421 from optical to TeV energies from February 13-17, 2010. 
Different markers denote the following dates: Square (gray)- SS data, Circle (black)- average of 2010 February 13-15 (low-flux state), Regular triangles (red)- 2010 February 16 (intermediate-flux state), Inverted triangles (blue)- 2010 February 17 (high-flux state). The spectral hardening can be seen in 
X-ray and TeV bands 
as the source 
became brighter.
The VERITAS data in high-flux state shows the highest flux state 
in TeV energies
during this giant outburst (see text for details). The EBL correction has been carried out according to the model presented in \citealt{domin}. The change in the slopes of lines joining optical and UV (see the inset plot) shows that the spectrum 
became harder
as the source 
brightened. 
The optical-UV data used in this work has been corrected for dereddening and host galaxy contribution.
}\label{fig:MWL}
\end{figure*}

Mrk 421
was detected during the 
flare, exhibiting
different flux states above 250 GeV from 2010 February 13-17 by
the HAGAR telescope array
located at Hanle, India (\citealt{HAGAR}). Activity of the source for these days was also reported by
the TACTIC 
telescope (in operation at Mount Abu in Western India) above 1 TeV (\citealt{TACTIC}; \citealt{TACTIC2}). 
Both HAGAR
and TACTIC observed Mrk 421 on 2010 February 16 and reported an enhancement in flux compared to previous days. 
The flux above 1 TeV observed by TACTIC on 2010 February 16 (MJD 55243.77758-55243.98686) (\citealt{TACTIC}) was (2.5$\pm$0.4) $\times10^{−11}$ ${ph}$ ${cm}^{−2}$ $s^{−1}$. The brightest flux state of Mrk 421 was observed by VERITAS on 2010 February 17 (55244.3-55244.5) above 100 GeV (see Figure \ref{fig:MWL} for details). The observed flux was 8 times 
that of the Crab.
On the same night HAGAR also reported the detection of Mrk 421 in 
the brightest state, showing
a flux of 6-7 times 
that of the Crab
above 250 GeV. 
However, TACTIC (\citealt{TACTIC2}) reported a lower flux on 2010 February 17 as compared to VERITAS (\citealt{VERITAS})
. From the MJDs quoted above it is clear that the observations were not strictly 
simultaneous,
and hence it is highly probable that on 2010 February 17, TACTIC caught only the decaying part of the flare. For our analysis, we have corrected for attenuation of 
the VHE spectrum
by the extragalactic background light (EBL) using the model of \citealt{domin}.\\

\begin{table*}
\centering
\begin{tabular}{r c c c c }
\hline
Day (MJD) & Optical & X-ray & $\gamma$-ray & VHE $\gamma$-ray\\\hline
2010 February 13-15 & Steward-Observatory,  & \textit{Swift}-XRT, -BAT&\textit{Fermi}-LAT  & HAGAR\\
(55240 to 55242) & \textit{Swift}-UVOT & & \\
2010 February 16 & Steward-Observatory, & \textit{Swift}-XRT, -BAT    &\textit{Fermi}-LAT  & HAGAR, TACTIC\\
(55243) & \textit{Swift}-UVOT & & \\
2010 February 17 & Steward-Observatory, & \textit{Swift}-XRT, -BAT    & \textit{Fermi}-LAT & HAGAR, VERITAS\\
(55244) & \textit{Swift}-UVOT & & \\
\hline
\end{tabular}
\caption{
Data collected from different instruments.
The optical V-band data from Steward observatory and UV data from \textit{Swift}-UVOT (using three filters UVW2, UVM2, UVW1) 
have been used.
The \textit{Fermi}-LAT data has been analyzed using latest Fermi tools (see texts for details). The X-ray data for different days has been used from \citealt{HAGAR}. For intermediate- and high-flux states the TeV data from TACTIC (\citealt{TACTIC}) and VERITAS (\citealt{VERITAS}) have been used, respectively.
In addition, we have used the available VHE data from HAGAR (\citealt{HAGAR}) for different states to approximate the flux level  at those energies and not for performing SED data-fitting.
}\label{tab:CollectedData}
\end{table*} 

\subsection{Summary of MWL data:}

It can be seen from the above discussion that Mrk 421 had undergone 
a change in flux
and spectral hardness from a low-soft state to a high-hard state in a matter of 5 days from 2010 February 13-17 in X-rays, GeV and TeV $\gamma$-rays. However, in the optical band there was no significant change in flux during that time. 
The inset
plot of Figure \ref{fig:MWL} shows that the slopes of the lines joining the optical and the X-ray spectrum in the SEDs 
became steeper
as the 
flare set
in during the low-flux state. This implies that the X-ray spectrum, 
in the 2-10 keV
energy range, became harder and the synchrotron peak shifted to higher energies in the hard X-rays as the flare progressed to the high-flux state. During this progression, a quasi-linearity was also observed between 1-10 keV X-rays and $\gamma$-rays above 100 GeV (\citealt{HAGAR}; \citealt{TACTIC}). Such strong variations in both X-ray and $\gamma$-rays and the quasi-linear correlation between the two suggests that the flare has a leptonic 
origin rather than
a hadronic 
one,
as no correlation between X-ray and TeV variabilities is expected under a hadronic scenario (\citealt{Aharonian:2000pv}; \citealt{2003APh....18..593M}). The details of the data used for this work are listed in Table \ref{tab:CollectedData}. As can be seen from the inset plot in Figure \ref{fig:MWL}, the synchrotron 
curvature decreased
as the flux 
increased
and the synchrotron peak 
shifted
to higher energies. This variation in the shape of the synchrotron peak could be associated with statistical or stochastic acceleration mechanisms (\citealt{2012MmSAI..83..122D}). As discussed in \S 4, the prevalence of these acceleration mechanisms during the flare is further supported by the value of electron energy indices that were obtained as best fit values for reproducing the three spectral states.
\section{Model description:}

We have used the MUlti-ZOne with Radiation Feedback (MUZORF) model of \citealt{Joshi2011} to individually reproduce the three flux states and the SS of Mrk 421, as 
presented
in \S 2.5, in a time-dependent manner. 
The model assumes a background thermal plasma of non-relativistic electrons, positrons, and protons in the jet. Two shells of plasma with different mass and velocity are assumed to be injected into the jet from the central engine that consists of a super-massive black hole (SMBH) and an accretion disk.
A merged shell is formed when a faster moving inner shell ejected at a later time catches up with a slower moving outer shell ejected at an earlier time. This merged shell serves as an emission region inside which two internal shocks, 
a reverse shock (RS) and a forward shock 
(FS), are produced.
In the frame of the shocked fluids, the shocks (FS/RS) travel in 
opposite
directions from each other and the corresponding emission regions are called the forward and reverse emission regions (\citealt{Joshi2011}). 
As the shocks propagate through their respective emission regions, a fraction of the internal kinetic energy of the shocked fluid that is stored in the baryons, is transferred to electrons and positrons. The particles in these regions subsequently get accelerated to relativistic and non-thermal energies due to this transfer.
The fraction of the bulk kinetic energy density of the shocked fluid that gets converted into magnetic and particle energy density is quantified by their respective partition parameters, $\varepsilon'_B$ and $\varepsilon'_e$. The ratio of these parameters provides a first-order estimate of the equipartition parameter, $e_B'$, of the model. We consider this as first-order because the time-dependence of our model makes the particle energy density evolve with time. Consequently, the equipartition parameter of our model is time-dependent. Hence, the above-mentioned ratio provides only an initial value of the equipartition parameter resulting from the model. 
The observed emission from the jet is a result of the energy lost by highly accelerated particles in the shocked medium. As the shocks move into their first radiating zone, they accelerate a fraction of its particle population, which subsequently loses energy 
via the synchrotron
and SSC mechanisms and produce the observed radiation. The model takes into account Klein-Nishina effects for calculating inverse Compton emission. 
Throughout the paper, we consider primed quantities representing co-moving or shocked plasma frame values, starred quantities stand for the observer's frame, and unprimed variables denote 
the rest frame of the host galaxy (lab frame).
The cylindrical emission region is considered to have a co-moving radius $R^{\prime}_{cyl}$ with a width $\Delta^{\prime}_{cyl}$. The radiation mechanism inside the entire cylindrical emission region is mainly governed by a randomly 
oriented magnetic field ($B^{\prime}$)
and the injected electron energy distribution (EED). 
The injected EED
follows a simple power law ($N(E) = N_0 E^{-q^{\prime}}$) with a single index $q^{\prime}$ and minimum and maximum electron energy distribution cutoffs represented by their corresponding Lorentz factors $\gamma^{\prime}_{min}$ $\&$ $\gamma^{\prime}_{max}$, respectively. 
The fraction of electrons, $\zeta'_e$, that get accelerated behind the shock fronts into a power-law distribution along with $\varepsilon_e'$ govern the initial value of $\gamma^{\prime}_{min}$ of the injected electron distribution.
Once the shocks exit their respective zones, $\gamma'_{min}$ $\&$ $\gamma'_{max}$ of those zones are made to evolve with time according to the formalism:
\begin{equation}
\gamma_{min(/max)}^{new} = \gamma_{min(/max)}^{old} - {\dot{\gamma}}  \big\vert_{min(/max)} dt
\end{equation}
, where\\ $dt$ is the time step of the simulation.\\

The following two equations govern the dynamics of particle and photon populations inside the emission regions in a time-dependent manner (\citealt{Joshi2011}).\\
\begin{equation}
\frac{\partial n_e(\gamma,t)}{\partial t} = -\frac{\partial}{\partial t}\bigg[{\bigg({\dv{Q}{t}}\bigg)}_{loss} n_e(\gamma,t)\bigg]+ Q_e (\gamma,t)-\dfrac{n_e(\gamma,t)}{t_{e,esc}}
\end{equation}
$\&$
\begin{equation}
\frac{\partial n_{ph}(\epsilon,t)}{\partial t} = {\dot{n}_{ph,em}(\epsilon,t)}-{\dot{n}_{ph,abs}(\epsilon,t)}-\dfrac{n_{ph}(\epsilon,t)}{t_{ph,esc}}
\end{equation}
, where\\

$n_e(\gamma,t)$ is the lepton density inside the system\\

$\big(\dv{Q}{t}\big)_{loss}$ is the loss rate of particles\\

$Q_e(\gamma,t)$ is the sum of injection of leptons and $\gamma-\gamma$ 
electron-positron pair production rate
.\\

$t_{e,esc}$ is the electron escape time scale.\\

${\dot{n}_{ph,em}(\epsilon,t)}, {\dot{n}_{ph,abs}(\epsilon,t)}$ is the photon emission and absorption rates, respectively.\\

$\epsilon = {h \nu}/{m_e c^2} $ 
is
the dimensionless photon energy.\\

$t_{ph,esc}$ is the angle and volume averaged photon escape time scale for a cylindrical emission region.\\

The model addresses inhomogeneity in photon and electron density inside the emission region by dividing the emission regions into multiple zones and considering 
radiative transfer
within each zone and in between the zones. The emission region has been split into multiple zones only along the width of the cylinder. A fraction of the resulting volume and angle averaged photon density from a zone is fed subsequently to the zones on its adjacent sides. The feedback is essentially the angle and volume averaged photon density escaping (in forward, backward and 
sideways
directions) from a cylindrical emission region. The total photon output from a zone consists of three components: synchrotron, SSC, and forward (feed-up) \& backward 
(feed-down)
feedback received from adjacent zones. The radiation feedback scheme also lets us calculate the SSC emission accurately as compared to models with homogeneous one-zone emission regions. This is especially important for appropriately 
reproducing the high-energy
component of TeV blazars (such as Mrk 421), 
in which SSC is likely an important, or even dominant, process for producing high-energy photons (\citealt{Joshi2011}).

The cylindrical emission region is divided into 100 zones, with 50 in the forward and 50 in the reverse emission region. The steps to solve the above 
two equations
in order to get the time averaged photon spectrum are discussed in detail in the paper mentioned above. 
The frequency range of $7.5 \times 10^6$ Hz  to $7.5  \times 10^{29}$ Hz has been selected for carrying out the simulations. The chosen range of the EED is 
from
1.01 to $10^{7}$ for each simulation. Both ranges are divided into 150 grid points.
\section{Results and Discussion:}

The SEDs presented in Figure \ref{fig:MWL} represent the temporal evolution of Mrk 421 over a period of 5 days from a low- to high-flux state in February 2010. In addition, a typical state of the source has been presented in the form of SS for carrying out a comparison of key physical parameters between flaring and non-flaring states and 
how they
contribute toward the evolution of a flare in a source like Mrk 421. As can be seen from the figure, the low-flux state is different in its spectral hardness and flux level in X-rays and $\gamma$-rays as compared to that of the SS. Hence, it can be identified with an initial phase, where the flare is beginning to 
build up.
During the course of the flaring event the source underwent significant changes in flux and spectral hardness across X-ray and $\gamma$-ray wavebands. Such a behavior allows us to study the daily evolution of the source at different energies. In order to explore the physical phenomena occurring inside the jet, we have used MUZORF to constrain the role of key physical parameters, such as the bulk kinetic luminosity, 
nature of the
particle population, and magnetic field strength present in the system.
We reproduced the SED of each of the four spectral states individually, in a time-dependent manner. The model reproduces observational signatures of blazars: instantaneous SEDs that are used to understand the dynamical evolution of radiation components and LCs that are used for deriving the time-averaged SED based on a selected time window from these variability profiles (\citealt{Joshi2011}). It is the time-averaged SED that is used for carrying out data-fitting for each spectral state by varying key physical parameter values for that state through an iterative process. For our analysis, the parameters of the successful fit for a particular spectral state served as guiding points for carrying out fitting efforts for the next state. The model, however, is not designed to carry forward the information of physical parameters from the previous state to the next in a self-consistent manner.

In order to execute our data-fitting method, we first estimated some model-independent parameters for each spectral state using the data from their respective SEDs (\citealt{2003ApJ...596..847B}; \citealt{1998ApJ...509..608T}). We have considered a minimum variability timescale
($t^*_{var}$)
of the order of 1 day. This is justified because a flux doubling time scale of $\sim 1$ day can be seen at GeV, X-ray, and TeV energies from the multi-wavelength LCs for this time period (\citealt{HAGAR}).
The variability timescale of the blazar allows us to 
put a constraint
on the length of the cylindrical emission region. This gives us,
$\Delta'_{cyl}$ = $c$ $D$ $t^*_{var}$/(1 + z). We can estimate the radius of the cylindrical emission region by assuming it to have the same volume as a spherical emission region of radius $\Delta'_{cyl}$. In that case, $(4/3)$ $\pi$ $\Delta'^3_{cyl}$ = $\pi$ $R'^2$ $\Delta'_{cyl}$. The value of $R'$ can be estimated as, $(2/\sqrt3)$ $\Delta'_{cyl}$. Considering $t_{var}$ as 1-day and Doppler factor as 40, the values for the width of the emission region and radius of the emission region turns out to be $\sim$ $1.4 \times 10^{17} cm$ \& $\sim$ $1.0 \times 10^{17} cm$, respectively. 
We note that our chosen value of Doppler factor is higher than the typical value of $\sim$ 21 that is used for Mrk 421 (\citealt{SteadyState}) but is not uncommon for what has been previously used for this source to satisfactorily reproduce its TeV data (\citealt{Chen2011}). A lower value of Doppler factor results in a SSC peak at lower energies and a softer TeV spectrum that is not in accordance with our observed data for the three states. Hence, we initialize the Doppler factor to a slightly higher value of 40.
The derived set of model-independent parameters is presented in Table \ref{tab:BaseModel}. These parameters served as initial input parameters for simulating the spectral states and were modified to obtain the successful fits for each of the states. These 
parameters were estimated
based on the identification of the peak flux and frequencies of the two SED bumps, spectral indices below and above the synchrotron peak, variability timescale of the observations, and the Doppler factor of the source (\citealt{1998ApJ...509..608T}). Since the observed data may not always allow a precise determination of these 
observables, the subsequent
model-independent parameter estimates may not be that accurate and may not necessarily provide a good spectral fit. Nevertheless, for modeling efforts they can still be used as reasonable starting points and the values of key physical parameters can be adjusted in subsequent runs to obtain a satisfactory spectral fit of a given spectral state.

\begin{figure*}
   \centering
   \includegraphics[width=.33\textwidth,angle=270]{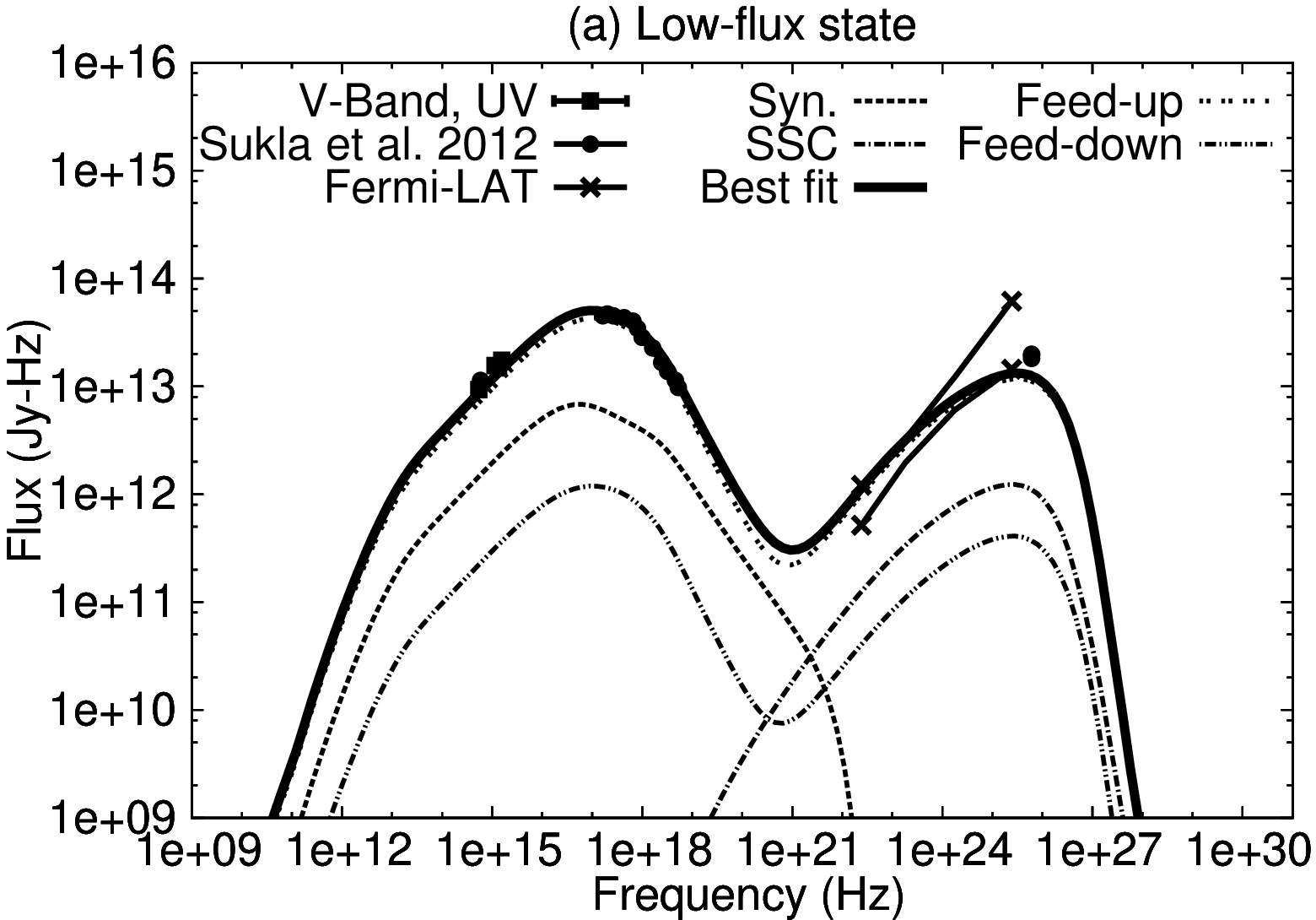}
   \includegraphics[width=.33\textwidth,angle=270]{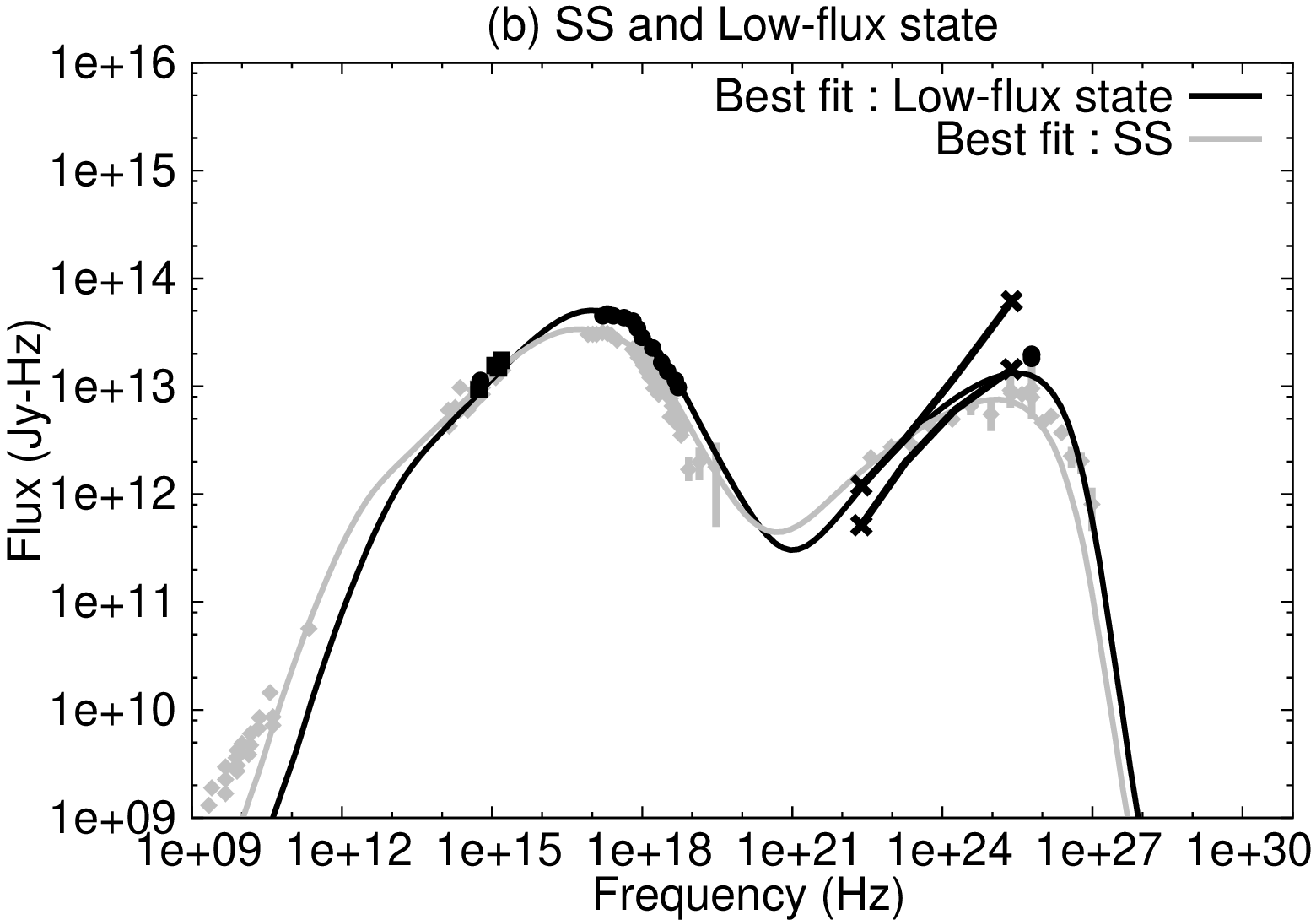}\\
   \includegraphics[width=.33\textwidth,angle=270]{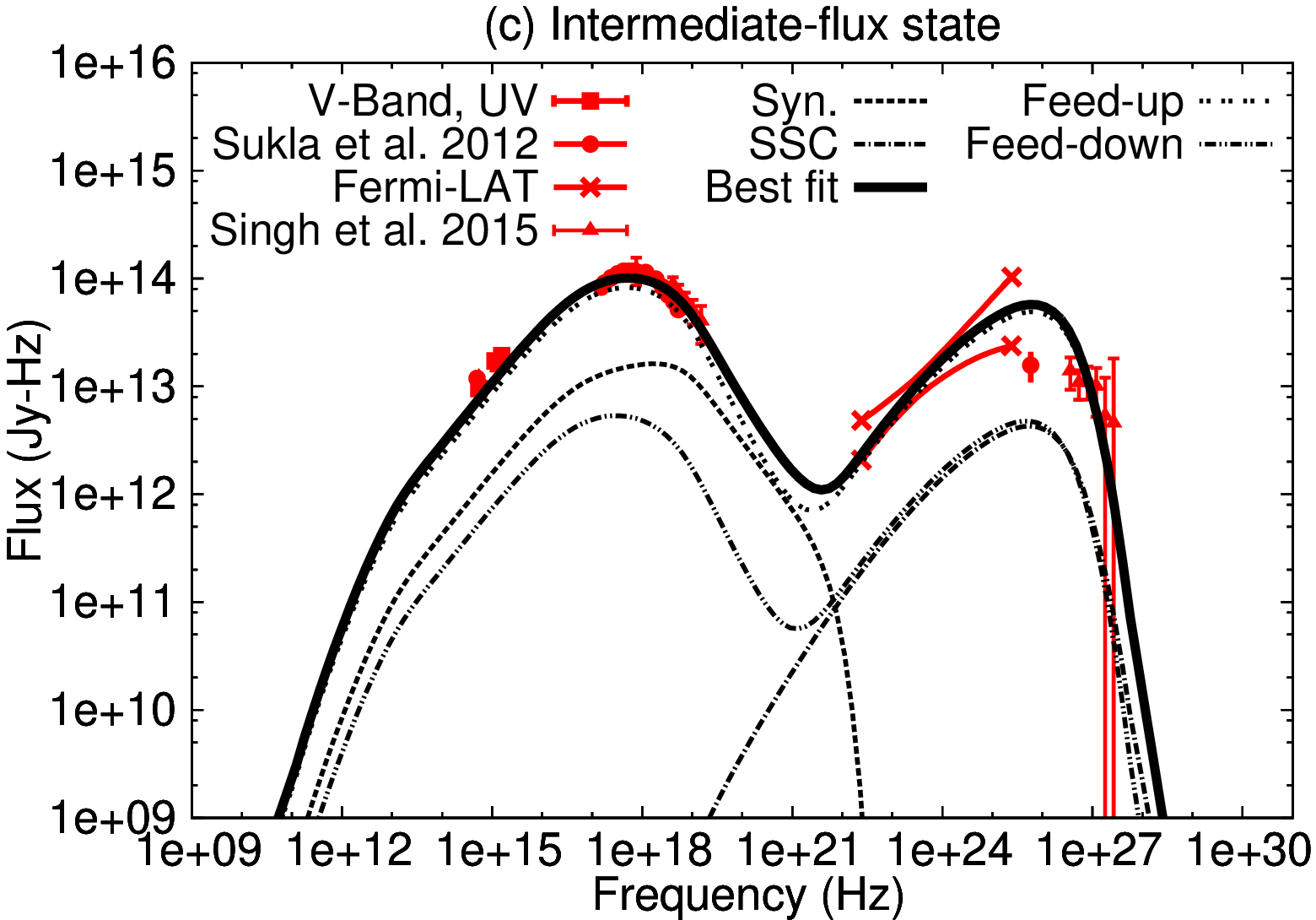}
   \includegraphics[width=.33\textwidth,angle=270]{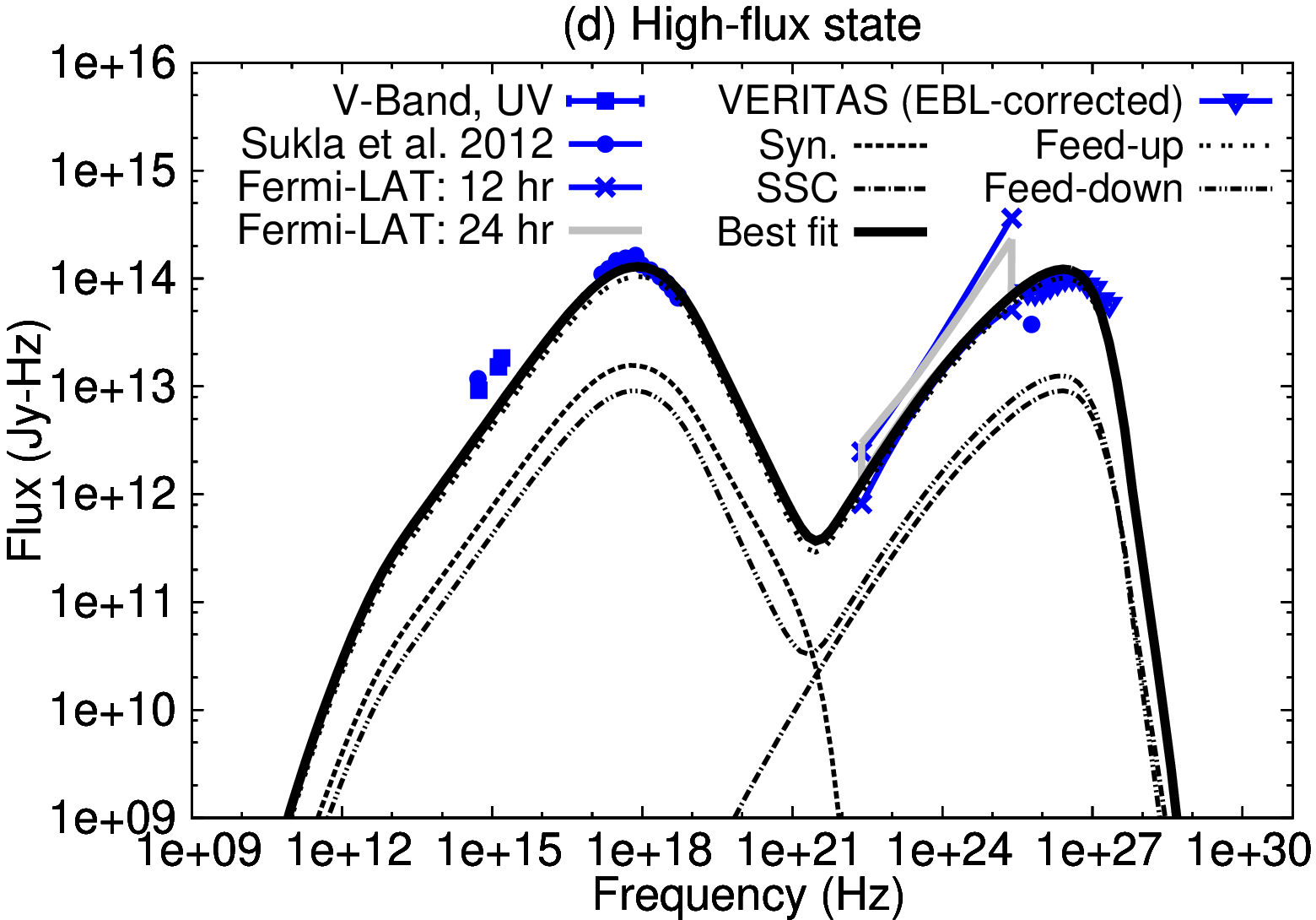}
\caption{The results of the SED modeling for different flux states: (a) low-flux state (2010 February 13-15), (b) steady state (4.5 month average as in \citealt{SteadyState}), (c) intermediate-flux state (2010 February 16), (d) high-flux state (2010 February 17). The black lines represent the successful models for each day. The {\it broken-line} and the {\it dash-dot} lines represent the synchrotron and SSC components in each of the successful models in (a), (c) and (d). The feed-up and feed-down components for (a), (c) and (d) are shown in {\it double dot-blank} and {\it dash-double dot} lines respectively. The butterfly plots for {\it Fermi}-LAT data on all the nights are drawn as solid lines with cross at the edges. The optical, X-ray and TeV data taken from \citealt{HAGAR} are shown by filled circles. The optical data from Steward Observatory and {\it Swift}-UVOT data are shown by square markers. The X-ray data and TeV data on 2010 February 16 (panel (c)) are from \citealt{TACTIC} and are shown by regular triangular markers. The TeV data from VERITAS on 2010 February 17 (panel (d)) is shown by inverted triangular markers. For details see text.}\label{fig:SEDs}
\end{figure*}

Close to one-thousand simulations were carried out in which we changed the value of key parameters iteratively.
We obtained a set of suitable parameters through this iterative process and chose successful models by visual inspection. As can be seen from Figure \ref{fig:SEDs}, this set of suitable parameters reproduces the observed data quite successfully.
The parameters that have been varied are the following: 
kinetic luminosity
($L_w$); inner and outer shell bulk Lorentz factors (BLF : $\Gamma^{\prime}_{i(0)}$); inner and outer shell widths ($\Delta^{\prime}_{i(o)}$); electron energy distribution parameter ($\varepsilon^{\prime}_e$); magnetic field ($B^{\prime}$); fraction of accelerated electrons ($\zeta^{\prime}_e$); electron injection index ($q^{\prime}$); zone(/jet) radius ($R^{\prime}_{cyl}$).
Among all the other parameters 
listed above, the kinetic
luminosity ($L_w$), width ($\Delta^{\prime}_{cyl}$) \& radius ($R_{cyl}^{\prime}$) of the emission region, and the electron injection index ($q^{\prime}$) have the maximum impact on the values of the magnetic field ($B^{\prime}$), $\gamma_{min}'$, and $\gamma_{max}'$.
\begin{table*}
\centering
\begin{tabular}{r c c c c c}
\hline
Parameter                     & Symbol                  & Steady state      & Low-flux state    & Intermediate-flux state & High-flux state \\  \hline
Doppler factor                & $D$                     & 40                & 40                & 40                      & 40                 \\
Slice/Jet Radius(cm)          & $R'_{cyl}$              & $1.0\times10^{17}$& $1.0\times10^{17}$& $1.0\times10^{17}$      & $1.0\times10^{17}$ \\
Shell Width(cm)               & $\Delta'_{cyl}$         & $1.4\times10^{17}$& $1.4\times10^{17}$& $1.4\times10^{17}$      & $1.4\times10^{17}$ \\
Equipartition parameter ($10^3$)
                              & $e_B'$       & 0.004               & 0.01               & 0.045                   & 1.0 \\
Particle Injection Index      & $q^{\prime}$            & 2.65              & 2.6               & 2.35                    & 2.15                \\
Magnetic field (mG)           & $B^{\prime}$            & 13                & 8                 & 4                       & 2                  \\
Minimum energy of an electron & $\gamma^{\prime}_{min}$ & $2.5\times10^{5}$ & $3.5\times10^{5}$ & $1.5\times10^{6}$       & $2.0\times10^{6}$  \\
Maximum energy of an electron & $\gamma^{\prime}_{max}$ & $1.3\times10^{7}$ & $3.2\times10^{7}$ & $9.0\times10^{7}$       & $1.1\times10^{8}$  \\
Variability timescale (s)     & $t^*_{var}$             & 1-day             & 1-day             & 1-day                   & 1-day              \\
\hline
\end{tabular}
\caption{
The list of initial input model parameters for different states. Different initial model parameters: $\Delta^{\prime}_{cyl}$, $R^{\prime}_{cyl}$, $q'$, $B'$, $\gamma_{min}$, and $\gamma_{max}$ are estimated from the data for different flux states, using a variability timescale of 1-day and a Doppler factor of 40, based on the prescription described in \citealt{2003ApJ...596..847B}. Here, we have used the optical to X-ray spectral index of Mrk 421 for various spectral states to estimate the injection index for our electron distribution. The magnetic field 
and electron energy density
values have been estimated according to the prescription given in \citealt{1998ApJ...509..608T} 
and \citealt{2008ApJ...686..181F}, respectively.
}\label{tab:BaseModel}
\end{table*}

\begin{table*}
\centering
\begin{tabular}{r c c c c c}
\hline
Parameter                  & Symbol          & Steady state                   & Low-flux state                 & Intermediate-flux state & High-flux State \\ \hline
Kinetic Luminosity (erg/s) & $L_w$           & $2.6\times10^{45}$             & $4.2\times10^{45}$ 	       &$9.5\times10^{45}$     & $9.8\times10^{45}$ \\
Doppler Factor             & $D$             & 40   			      & 40                 	       & 40                    & 44.0               \\
Bulk Lorentz Factor        & $\Gamma^{\prime}_{sh}$& 28                          & 28                             & 28                    & 40.0               \\
Magnetic Field (mG)        &  $B^{\prime}$   & 89 & 64                	       & 60                    & 26                 \\
Total shell Width (in cm)  & $\Delta'_{cyl}$ & $1.1\times 10^{17}$            & $1.1\times 10^{17}$ 	       & $1.1\times 10^{17}$   &$1.1\times 10^{17}$  \\
Equipartition parameter ($10^3$)  & $e_B'$   & 0.18              & 0.28                           & 1                     & 5.6\\
Fraction of accelerated electrons ($10^{-3}$)& $\zeta_e'$ & 37 & 4.5                            & 4.5                   & 3.7 \\
Min. energy of an electron &$\gamma^{\prime}_{min}$& $4.4\times 10^{2}$    & $1.2\times10^{3}$              & $1.1\times10^{3}$     &$1.2\times10^{3}$\\
Max. energy of an electron &$\gamma^{\prime}_{max}$& $3.1\times 10^{5}$    &  $4.0\times10^{5}$             & $7.1\times10^{5}$     &$9.5\times10^{5}$\\
Particle Injection Index   & $q^{\prime}$    & 2.1                         & 2.01                           & 1.86                  & 1.75               \\
Zone(/Jet) Radius (in cm)                    & $R_{cyl}'$         & $1.25\times 10^{17}$           & $1.25\times 10^{17}$           & $1.25\times 10^{17}$  &$1.25\times 10^{17}$ \\
Observing Angle(deg)                         & $\theta_{obs}^{*}$ & 1.3                            & 1.3                   & 1.3                & 1.3\\
\hline
\end{tabular}
\caption{The successful model parameters for different days of the flaring episode. A detailed descripton of the comparisons is presented in \S 4.}\label{tab:BestParam}
\end{table*}

\begin{figure*}
   \centering
   \includegraphics[width=.4\textwidth,angle=270]{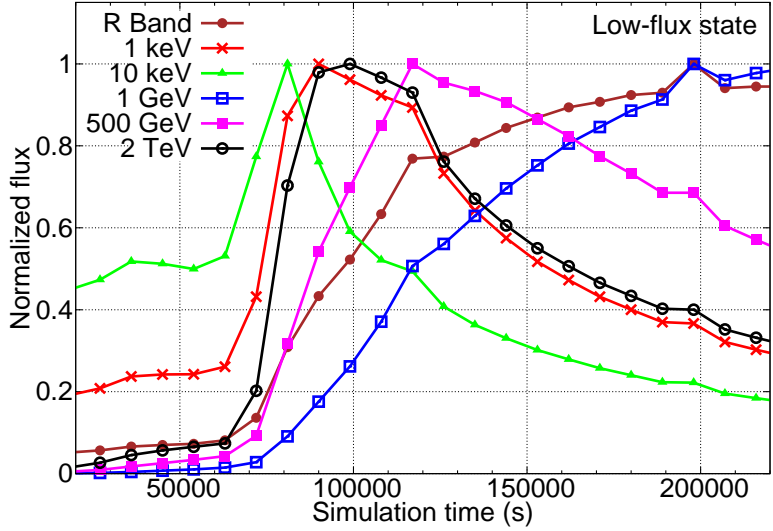}
   \caption{The simulated LCs (normalized to the peak flux for each of the LCs) of Mrk 421 at six different energy bands 
   (R-Band [brown-filled circle]; 1 keV [red-cross]; 10 keV [green-filled triangle]; 1 GeV [blue-hollow square]; 500 GeV [pink-filled square]; 2 TeV [black-hollow circle]) corresponding to the successful SED model for the low-flux state. The variation of flux rise time and flux decay time for 
   each energy band can be seen. The selection of time window was made such that both rise and decay of the LCs are considered inside the selected time window for making the time averaged SEDs. The LCs, except for the Optical R-Band, 1 GeV, and 500 GeV, peak around 80 ks. In case of the optical R-Band and 1 GeV, the radiation profiles are produced via relatively low-energy electrons. Hence the optical and GeV components peak outside the selected time window mentioned above.}
   \label{fig:lowstate}
\end{figure*}

\begin{figure*}
   \centering
   \includegraphics[width=.33\textwidth,angle=270]{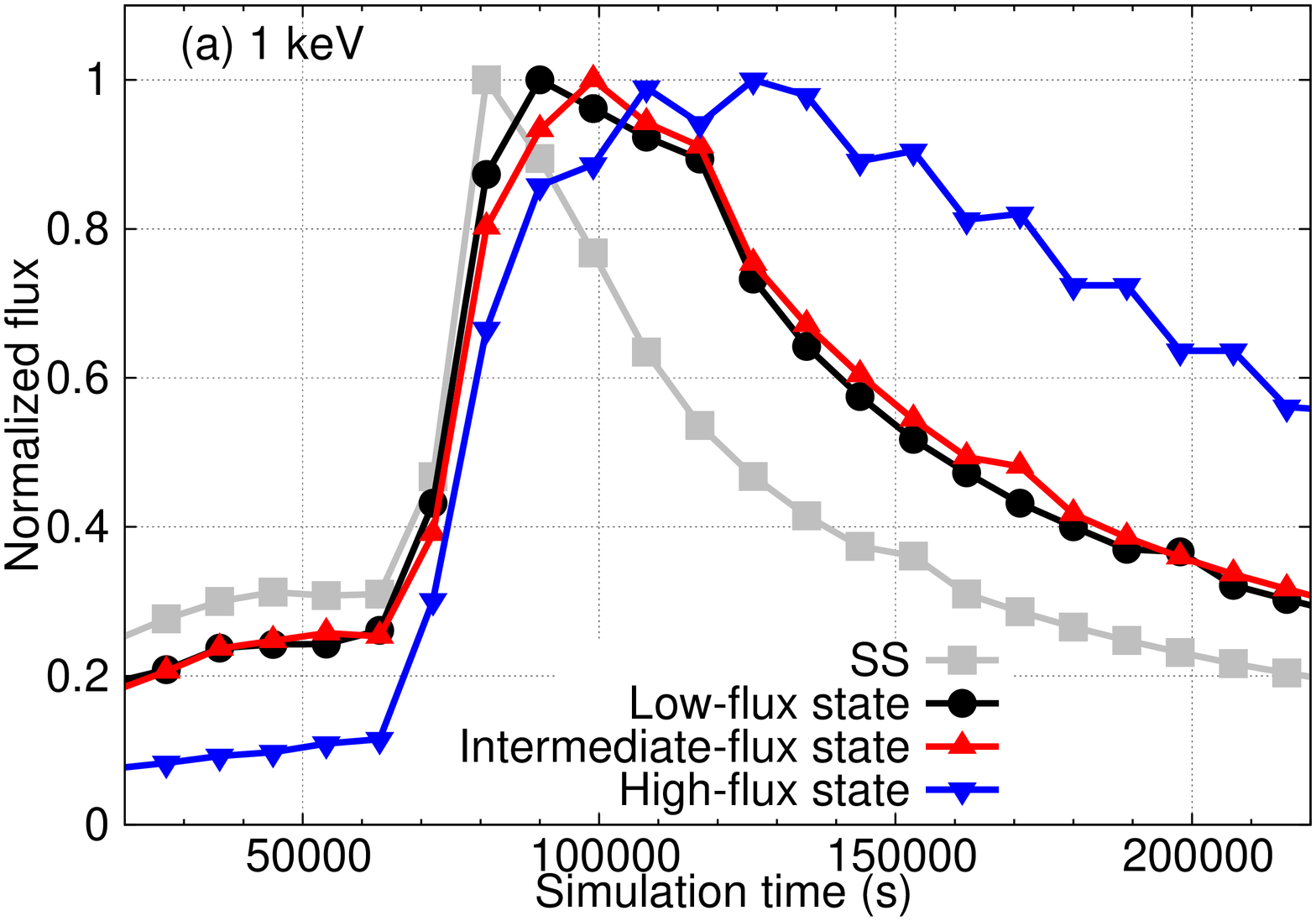}
   \includegraphics[width=.33\textwidth,angle=270]{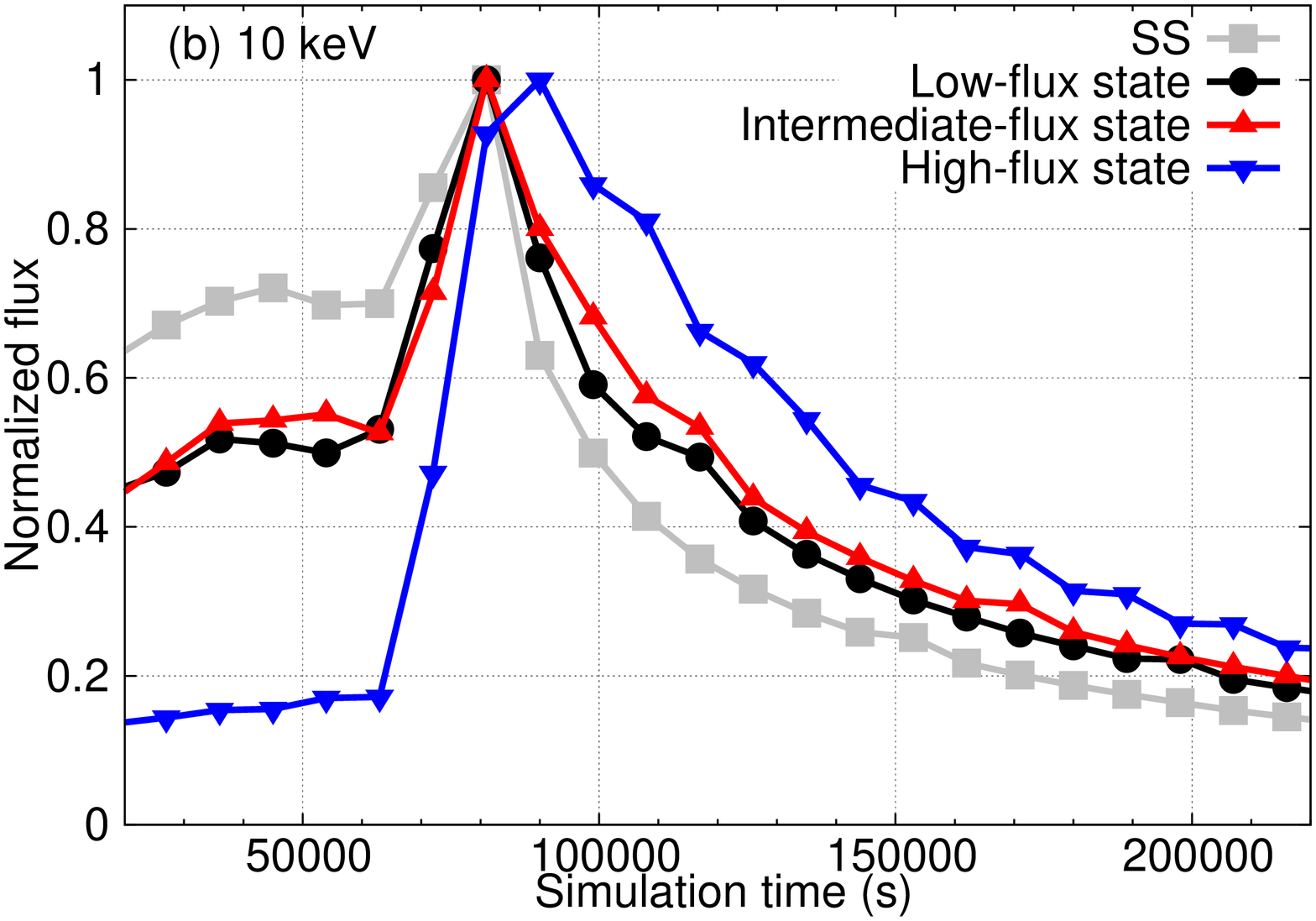}\\
   \includegraphics[width=.33\textwidth,angle=270]{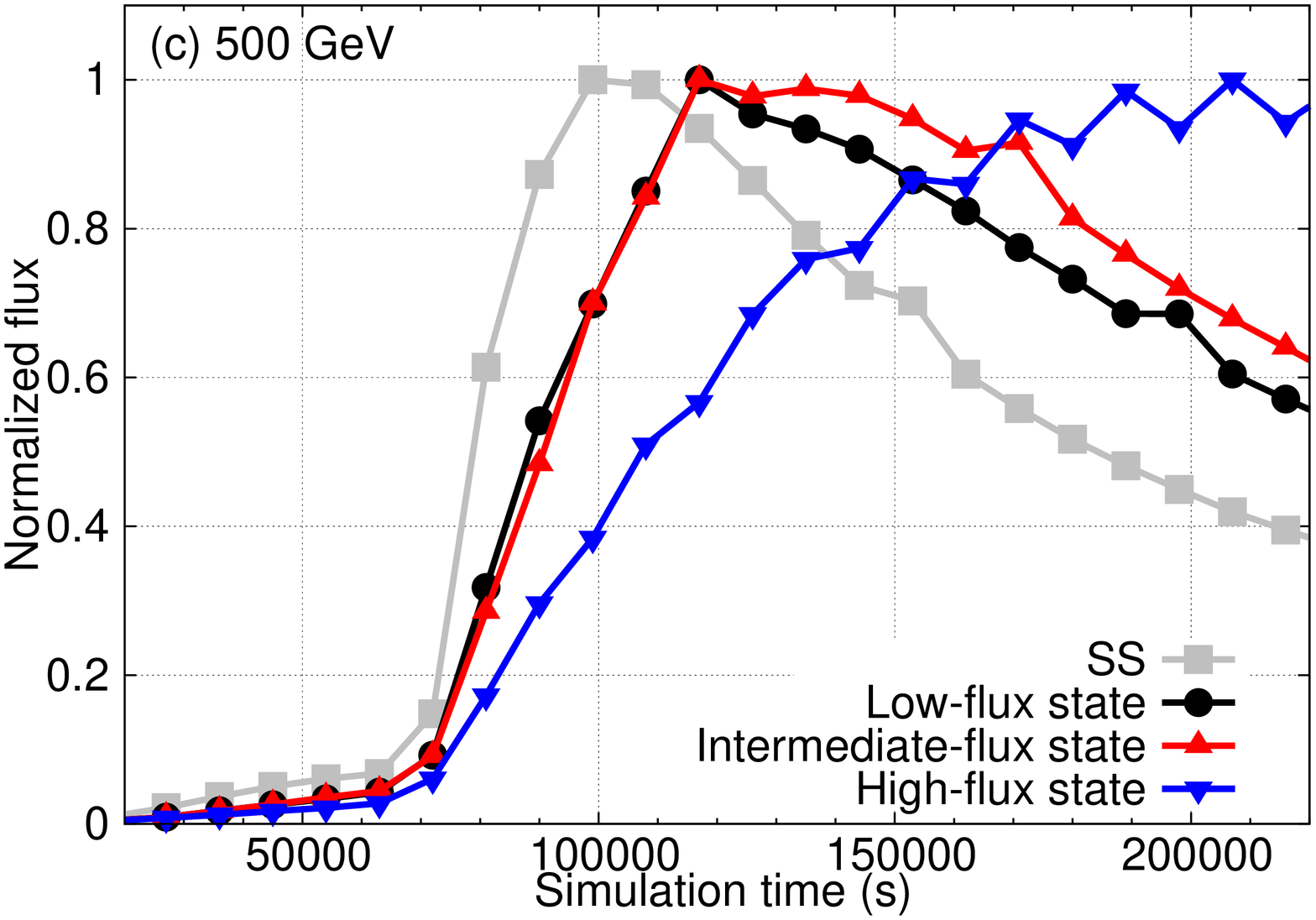}   
   \includegraphics[width=.33\textwidth,angle=270]{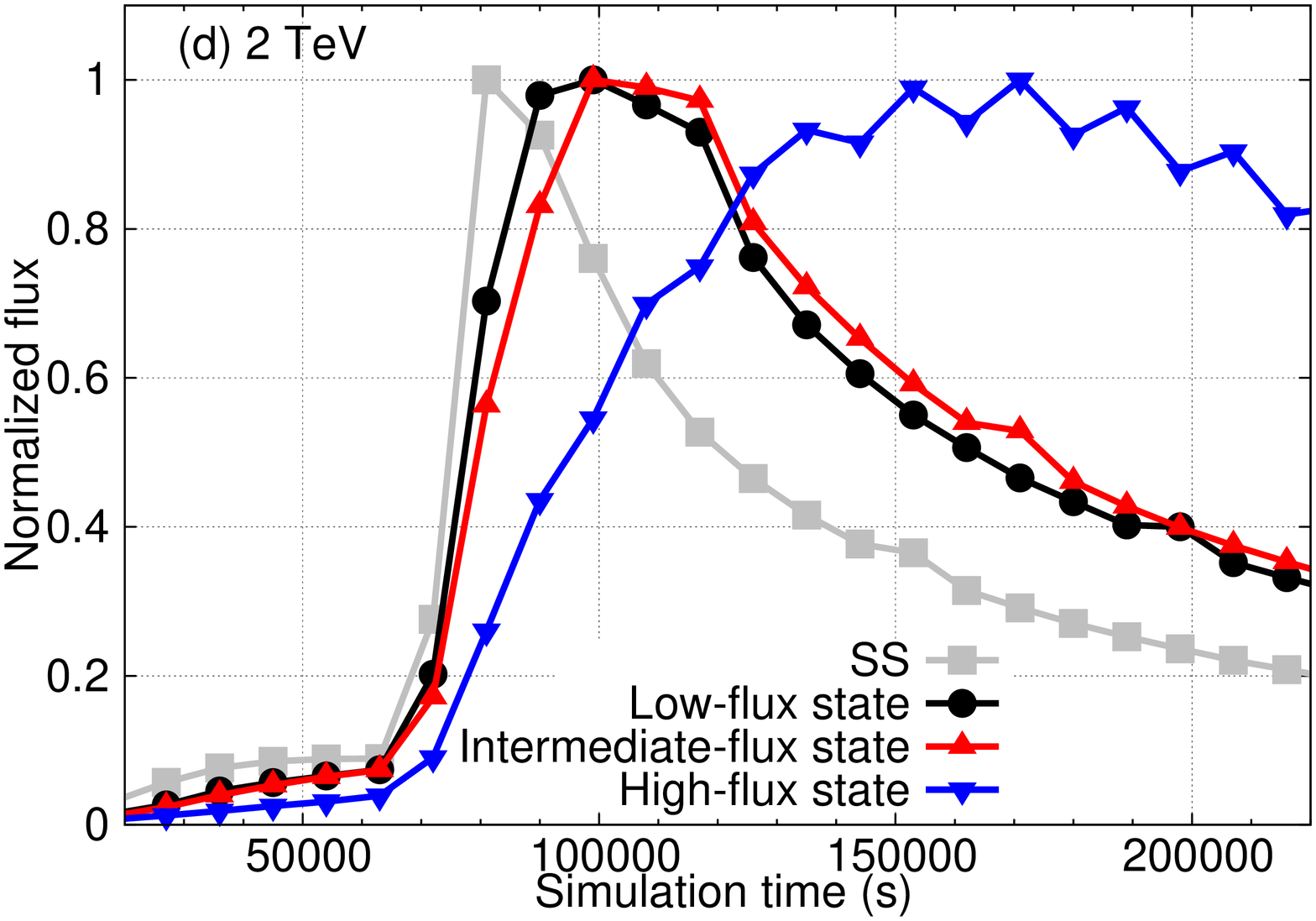}
   \caption{The normalized LCs (normalized to the peak flux value for each of the LCs) obtained from the model parameters in Table \ref{tab:BestParam}. Comparisons of simulated LCs for 1 keV, 10 keV, 500 GeV, and 2 TeV for four different spectral states: i) SS [gray-square], ii) Low-flux state [black-circle], iii) Intermediate-flux state [red-regular triangle], and iv) High-flux state [blue-inverted triangle] are shown in panel (a), (b), (c), and (d) respectively. The detailed description is presented in \S 4.}
   \label{fig:LCs}
\end{figure*}

We consider the mass of the SMBH to be  $M = 2 \times 10^{8}$ \(M_\odot\)
(\citealt{SteadyState}; $ M_\odot $: 1 solar mass) which predicts the Eddington luminosity as $2.6 \times 10^{46}$ erg/s. The values for 
the kinetic luminosity obtained from the model fit (see Table \ref{tab:BestParam}) for all the spectral states are lower than the value mentioned 
above, implying
that the states are sub-Eddington. 
Figure \ref{fig:SEDs} shows the result of our SED modeling for the four states considered here. The low-flux, intermediate, and high-flux states are shown in panels (a), (c) and (d) respectively. In panel (b), we show a comparison of the SED of the low-flux state with the SS of the source. As can be seen from various panels in the figure, the resultant time-averaged SEDs reproduce these spectral states satisfactorily.

The successful model parameters for all spectral states are presented in Table \ref{tab:BestParam}. 
As can be seen from Table \ref{tab:BaseModel}, the values of the initial input parameter set do not match the values obtained from successful fits of the four spectral states. 
However, the pattern of the values of the magnetic field strength, the equipartition parameter, and the particle injection does so.
The values of these parameters 
in Table \ref{tab:BestParam}
hint 
toward the existence
of a general trend 
in the role
of key physical parameters during the evolution of the flare over a period of three days. As the flare sets into the system 
during the low-flux
state and then evolves into the high-flux state, the kinetic luminosity that is injected into the system increases while the magnetic field strength
and the fraction of accelerated electrons decrease. 
This happens because as the flare evolves and higher energy bands become brighter and 
harder,
the corresponding synchrotron component needs to be suppressed sufficiently for the SSC component to rise up and attain its highest flux level in the high-flux state. This is further supported by an increased bulk speed of the emission region for the high-flux state and a more energetic population of particles with a harder distribution.

We note that the values of beaming parameters obtained in Table \ref{tab:BestParam} are fully consistent with the results of an independent study, not based on SED modeling, performed on probing the X-ray signature of recollimation shocks in \citealt{2019arXiv190406802H}. 
 The authors found the width of the emitting region (in the plasma frame) to lie within $0.43-19 \times 10^{17}$ cm, which is also in agreement with the value of total shell width listed in Table \ref{tab:BestParam}.

The successful fits for the four spectral states result in a particle injection index value that varies from $q’$ = 2.2 to 1.8. In general, relativistic shocks (with their shock normal parallel to the jet axis) are expected to produce a particle distribution with 2.2 < $q’$ < 2.3 (\citealt{Achterberg}; \citealt{Gallant}) through the Fermi first-order acceleration mechanism. On the other hand, indices with $q’$ < 2.0 could be produced from stochastic acceleration in resonance with plasma wave turbulence behind the relativistic shock front (\citealt{2004A&A...414..463V}; \citealt{0004-637X-621-1-313}). This implies that typically it is Fermi first order that is prevalent as the dominant acceleration mechanism in Mrk 421. However during the onset of a flare and course of its evolution, the dominant acceleration mechanism, within the shock model, changes from Fermi first to second order or stochastic acceleration. 
From Table \ref{tab:BestParam}, one can also see that successful model parameters indicate a departure from equipartition for all spectral states and a matter-dominated jet. This kind of a departure has previously been seen for Mrk 421 (\citealt{Sinha:2016pdu}; \citealt{HAGAR}). For our case, this implies that the shocks are mediated by the acceleration of particles and 
resulting in the energy density
of the particles to be much higher than that of the field.

In the resultant time-averaged SED (Figure \ref{fig:SEDs}(a)) for the low-flux state, the synchrotron hump  peaks in the soft X-ray at $\nu^{*}_{syn}$ = $1 \times 10^{17}$ Hz. The frequency at which we 
observe the SSC
peak is at $\nu^{*}_{SSC}$= $1.5 \times 10^{25} Hz$. The spectral upturn can be seen at around $\nu^{*}_{turn}= 3.5 \times 10^{20}$ Hz. This is the frequency 
above which
the spectrum becomes SSC dominated. The position of the spectral upturn indicates that the synchrotron photons extend into hard X-rays. Figure \ref{fig:SEDs}(a) shows that the optical R-band and 1 keV photons are purely synchrotron dominated while the radiation at 10 keV is mostly produced due to synchrotron emission with small contributions from SSC. 
On the other hand, photons at energies 1 GeV, 500 GeV, and 2 TeV are purely SSC-dominated. The Compton dominance factor (CD), defined as the ratio of the peak flux of SSC hump ($\nu F_{\nu}^{* SSC, peak}$) to that of the synchrotron hump ($\nu F_{\nu}^{* Syn., peak})$ for this simulation is ($\nu F_{\nu}^{* SSC, peak}$/$\nu F_{\nu}^{* Syn., peak}$=) 0.29. Figure \ref{fig:SEDs}(b) shows a comparison between the models for SS and the low-flux state. As can be seen from the figure, the synchrotron and SSC peaks for the SS appear at almost the same frequencies as that of the low-flux state. However, the overall flux level is slightly less compared to the low-flux state (see Figure \ref{fig:SEDs}(b)). In order to reproduce a lower SSC contribution in the SED of SS, a slightly higher value of $B^{'}$ is required in accordance with the general trend of key parameters.\\
The model parameters (shown in Table \ref{tab:BestParam}) for the successful SEDs for SS and the low-flux state support our assumption that the low-flux state carries signatures of the initial phase of the flare.\\
The SED that closely fits the data of the intermediate-flux state is shown in Figure \ref{fig:SEDs}(c) and the corresponding model parameters are given in Table \ref{tab:BestParam}. 
In order to fit the SED corresponding to this 
spectral state, we
decreased the value of $B^{\prime}$, which increased $\gamma^{\prime}_{max}$ and the Compton dominance of the system. In order to correct for this effect on the overall SED, we increased the value of $L_w$  and also decreased the 
value of $q^{\prime}$, which 
resulted in bringing the flux of both synchrotron and SSC components to the right level.
We note that our model for the intermediate- and high-flux states overproduce the HAGAR data points located at $\sim$ 250 GeV. In the absence of any data from imaging telescopes, HAGAR data point was used as a guide to approximate the flux level at those gamma-ray energies and was not meant to be used for performing data fitting. This is because HAGAR uses wave-front sampling technique\footnote{http://www.tifr.res.in/$\sim$hagar/telescopes$\_$details.html} that is known to have systematic uncertainties in the estimation of flux much larger than that of the imaging telescopes operating at TeV energies, such as VERITAS and TACTIC.

The source was observed in the brightest state with highest $\gamma$-ray flux on 2010 February 17. The successful model for this SED is shown in Figure \ref{fig:SEDs}(d) and values of the corresponding key physical parameters are listed in Table \ref{tab:BestParam}.
For this day the SSC component peaked at around 2 TeV. This SSC peak position is at a relatively higher energy as compared to other days and is indicative of spectral hardening of the source in this energy regime. 
The CD for the high-flux state is 0.85, indicating an increase in the peak of the SSC component as compared to the low-flux state. 
The magnetic field value has been decreased further to 26 mG in order to obtain a particle population with a slightly higher $\gamma_{max}^{\prime}$ in comparison to its low-flux state counterpart. In addition, a higher value of $\Gamma^{\prime}_{sh}$ and 
therefore
$D$ shift the SED toward higher energies. An even harder value of $q^{\prime}$ = 1.75 has been used to obtain the successful model for this state as compared to other states under consideration. The recession of $q^{\prime}$ to lower values, as the source evolves from a low-flux state to the brightest state, imply an interplay of Fermi first- and second-order acceleration mechanisms. Such an interplay along with contribution from shear acceleration has been suggested to energize particles to higher energies (\citealt{Rieger}).\\
As can be seen from Figure \ref{fig:SEDs}(d), the overall optical flux for the high-flux state is slightly 
under-produced.
This is mainly because as we go from low- to high-flux 
state, we
need lower magnetic field in order to fit the observed SSC component. A low magnetic field results in low synchrotron 
output, which
is reflected in the optical region of the SED. 
Besides, the high-flux state is mostly devoid of low-energy 
electrons, as
the corresponding particle population is much harder and more energetic compared to the other two states. As a result, it is not possible to reproduce optical emission for that day successfully. In addition, the source did not exhibit any variability in the optical energies over the course of this flaring event. Since the emission at optical and GeV-TeV energies are not correlated 
at all, it is
possible that the optical emission for the high-flux state 
originates
from a slightly larger emitting volume than what has been considered for this day.

Figure \ref{fig:lowstate} shows the LC 
profiles of the
low-flux state for six different wavebands corresponding to the successful fit shown in Figure \ref{fig:SEDs}(a). The LC profiles for all days have been evaluated at frequencies: R-Band [$4.68 \times 10^{14}$ Hz]; 1 keV [$2.42 \times 10^{16}$ Hz]; 10 keV [$2.42 \times 10^{17}$ Hz]; 1 GeV [$2.42 \times 10^{23}$ Hz]; 500 GeV [$1.21 \times 10^{26}$ Hz] and 2 TeV [$4.84 \times 10^{26}$ Hz]. We calculated LCs at these energies because most of the current generation telescopes are most sensitive in these energy bands. These are also the representative LCs for synchrotron (optical and 1 keV LCs) and SSC (2 TeV LC) components for typical HBLs, such as Mrk 421. All the different LCs  are characterized by different flux rise times (time scale associated with flux reaching the peak value) and flux decay times (time scale associated with flux decaying from the peak value). In order to calculate time-averaged SEDs for various spectral states appropriately, we used the corresponding simulated LCs and selected the time window of $\sim$ 90 ks in such a way that both the rising and decaying components of the LCs are properly taken into consideration (see Figure \ref{fig:LCs}). 
The fact that the peak of the SSC component appears at $\sim$ 2 TeV for the high-flux state and a harder population of electrons is required to reproduce this feature implies that the pulse at 2 TeV now receives contribution from relatively lower energy electrons as compared to its counterparts for the other three states. As a result, the pulse decays slowly and lasts in the system for a much longer duration. In order to cover the entire flare, we have considered a time-window from 60 ks to 160 ks for the low and intermediate-flux state, whereas, for the high flux state a time-window of 65 ks to 195 ks has been selected.

As shown in the figure, the flare at 10 keV is the first to peak because it is a result of synchrotron emission from high-energy electrons. The peak positions of synchrotron-dominated flares (1 keV \& R-Band) shift to later times depending on the energy of electrons responsible for producing them. The SSC-dominated flares are expected to lag behind their synchrotron counterparts. Hence, all $\gamma$-ray flares peak later. However, depending on the energy of electrons responsible for producing SSC emission at 
these wavebands,
the high-energy SSC flares (2 TeV \& 500 GeV) would peak sooner than the low-energy ones (1 GeV). This gradation in peaking times due to the difference in cooling times of high- and low-energy electrons can be easily seen in the LC profiles of the low-flux state in Figure \ref{fig:lowstate}. The profiles also show a strong correlation 
between the synchrotron-dominated
R-band flare and SSC-dominated 1 GeV flare. The optical flare is a result of synchrotron emission from low-energy electrons compared to those responsible for the synchrotron emission in the X-rays. Hence, the corresponding flare peaks at a later time and lasts in the system for a longer time. In addition, a fraction of these R-band photons is also responsible for producing 1 GeV photons 
via the SSC
mechanism through low-energy electrons. Hence, the corresponding flare is strongly correlated with the optical one. On the other hand, the 500 GeV SSC-dominated flare 
gets contributions
from both X-ray and low-energy photons. As a result, it exhibits a mild correlation with the X-ray flare. 

\begin{figure*}%
    \centering
    \subfloat[][]{\includegraphics[width=.35\textwidth,angle=0]{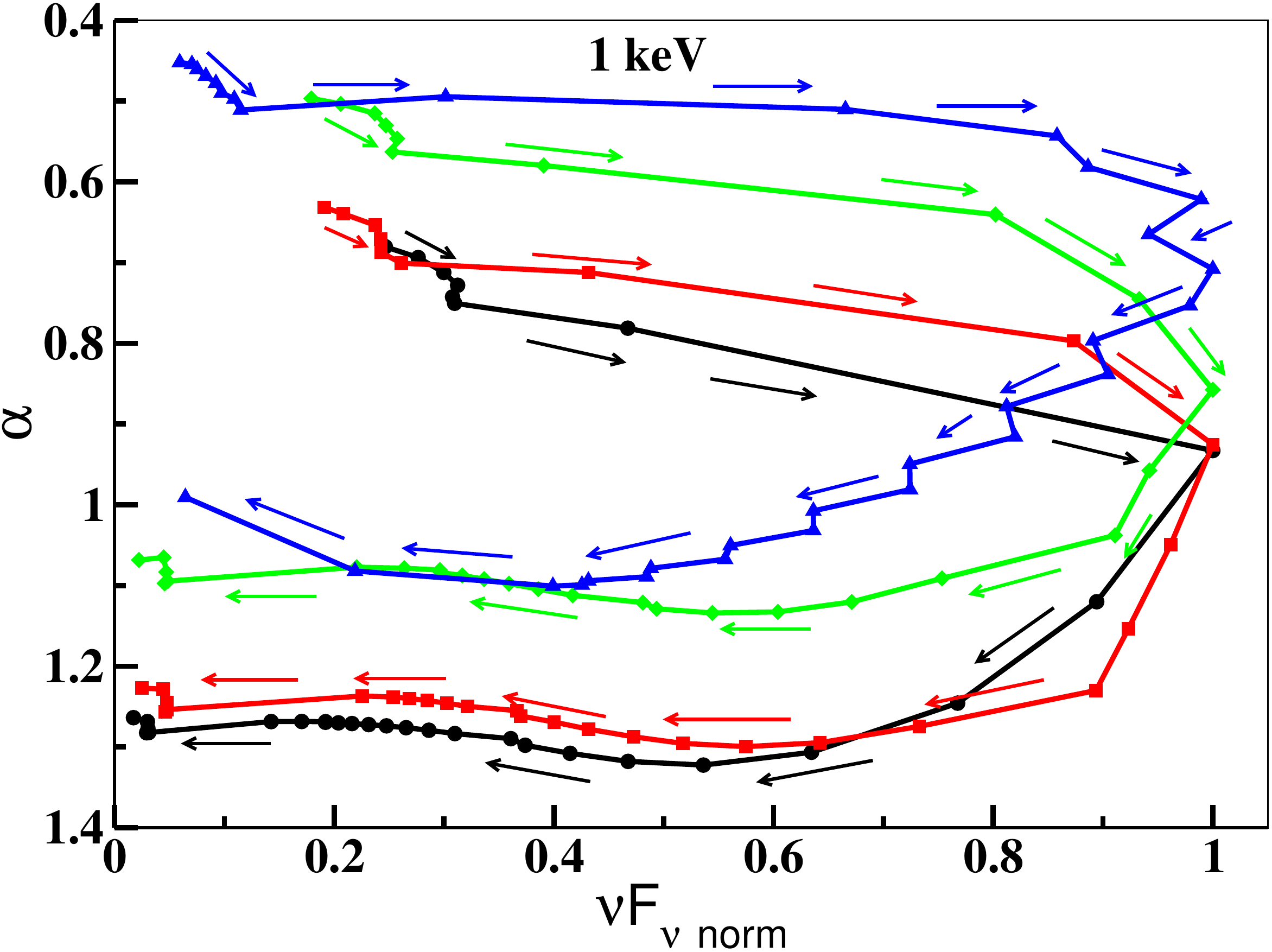}}\quad%
    \subfloat[][]{\includegraphics[width=.35\textwidth,angle=0]{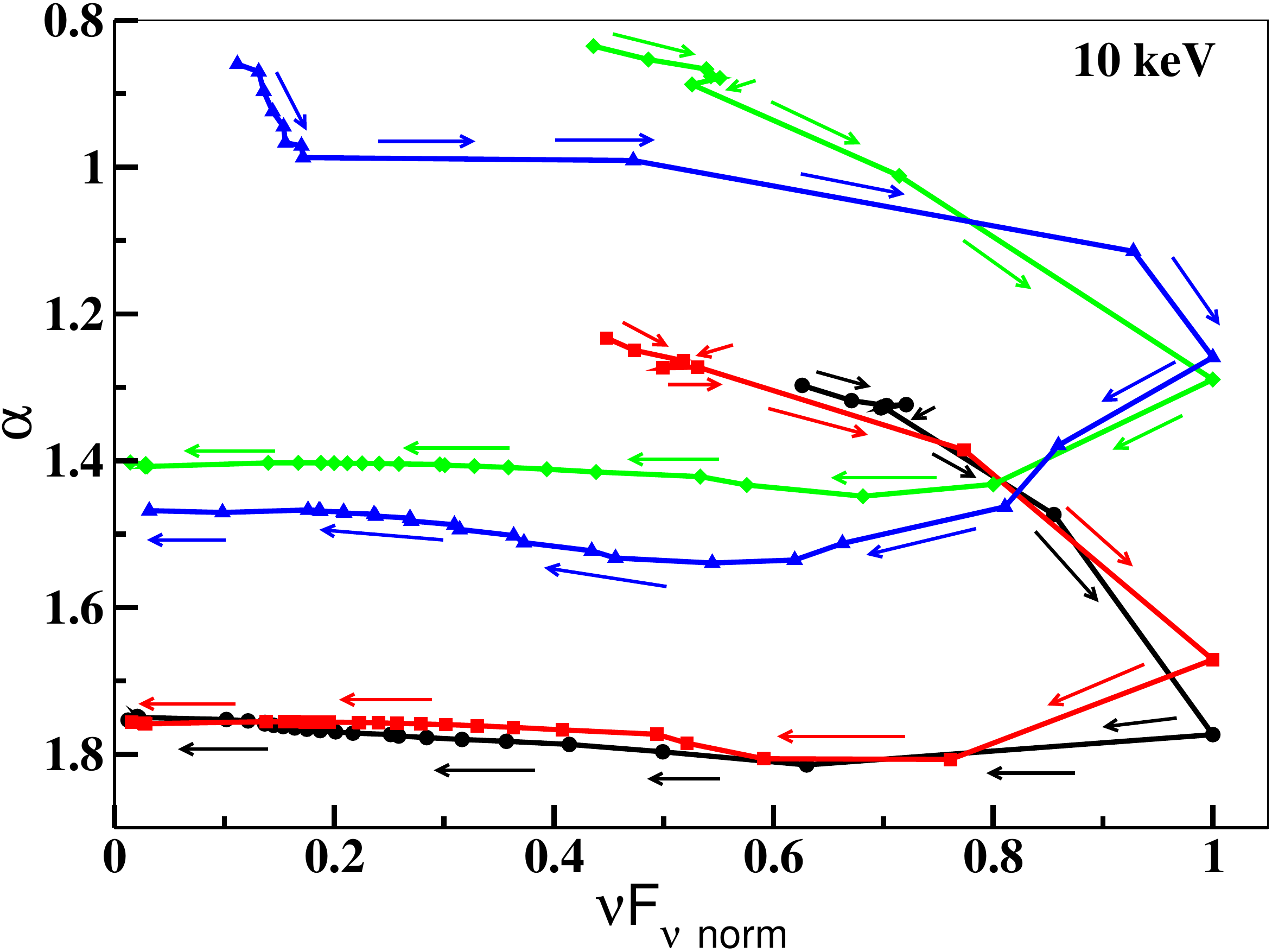}}\\%
    \subfloat[][]{\includegraphics[width=.35\textwidth,angle=0]{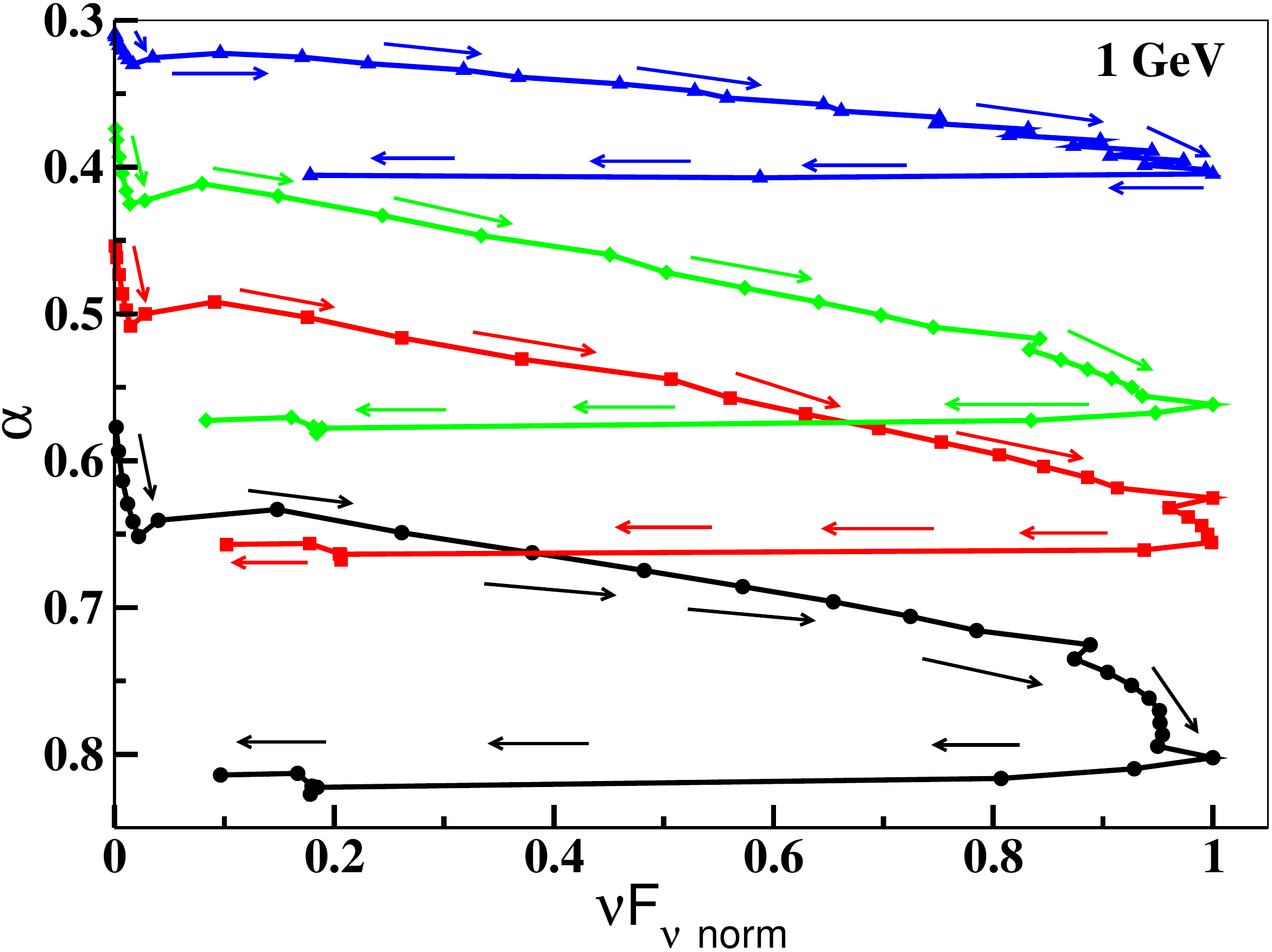}}\quad%
    \subfloat[][]{\includegraphics[width=.35\textwidth,angle=0]{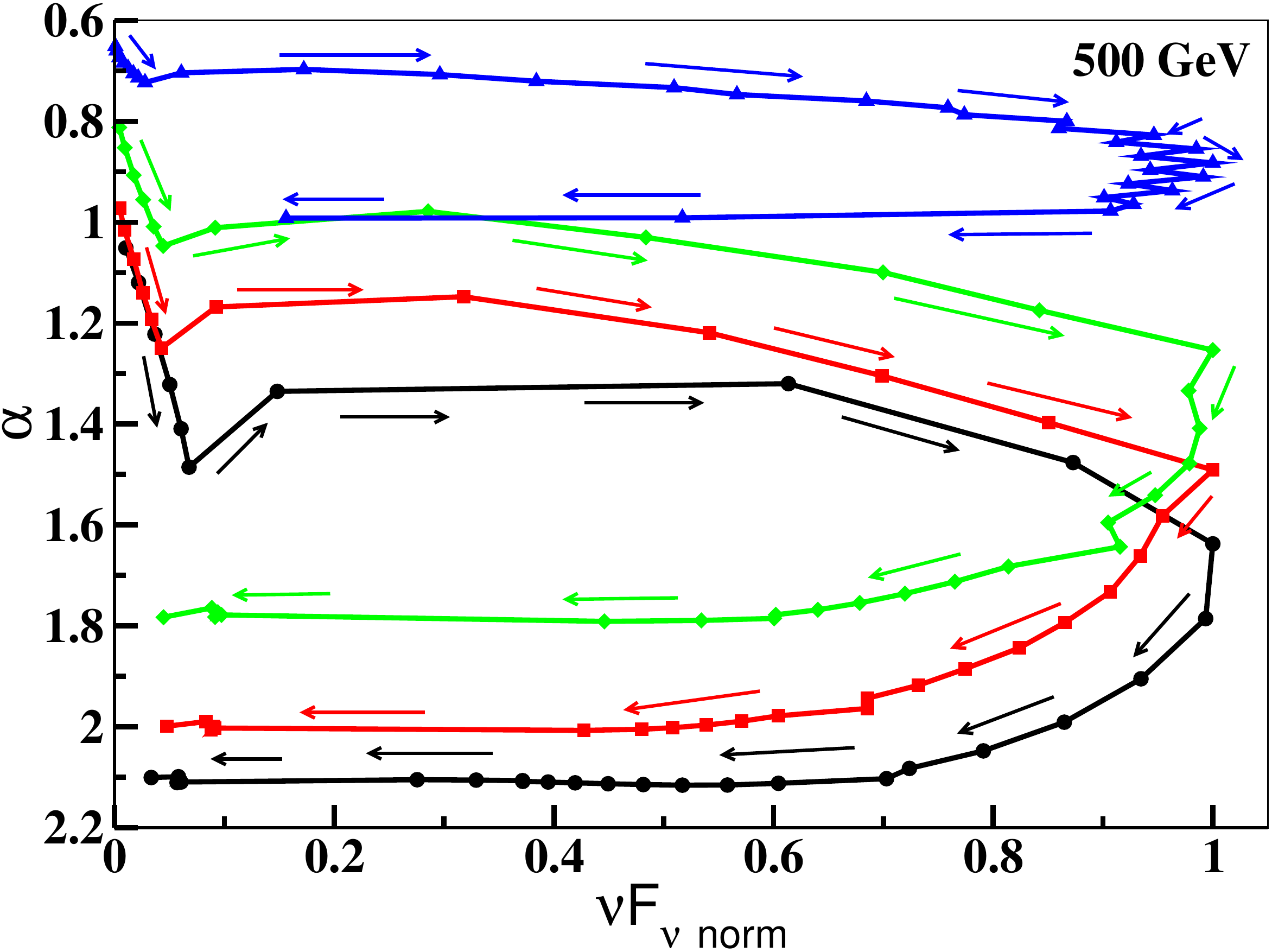}}\\%
    \caption{Spectral evolution of Mrk 421 as a function of flux calculated at X-ray (1 \& 10 keV) and $\gamma$-ray (1 \& 500 GeV) energies for the four spectral states, SS (solid black line with filled circles), low-flux state (solid red line with filled squares), intermediate-flux state (solid green line with filled diamonds), and high-flux state (solid blue line with filled triangle-up): a) top-left panel shows spectral evolution at 1 keV; (b) top-right panel shows the same for 10 keV; (c) bottom-left shows for 1 GeV; and (d) bottom-right shows the evolution calculated at 500 GeV. The source shows a clockwise pattern in all cases indicating that softer energies lag behind the harder ones. The photon spectral index at each of the energies is denoted by $\alpha$ and follows the relationship, $F_{\rm \nu} \propto \nu^{-\alpha}$.}
   \label{fig:hyst}
\end{figure*}

The above-mentioned behavior of flares at various energy bands was present for all days under consideration and is typical of what has been previously seen in Mrk 421. Panels a, b, c, \& d of Figure \ref{fig:LCs} show a comparison of the LCs, normalized to their 
highest
flux values, as predicted by successful models for those days at 1 keV, 10 keV, 500 GeV, \& 2 TeV, respectively. The LCs at optical and 1 GeV are excluded from the 
figure, as these bands
did not show any significant variation in their profiles during the evolution of the flare. This is also consistent with the Mrk 421 optical LC presented in \citealt{HAGAR} for this time period. Table \ref{tab:BestParam} shows that the successful model for 
intermediate-flux state
requires a higher kinetic luminosity, lower magnetic field and harder injection index as compared to that for the low-flux state. This resulted in a larger Larmor radius and a higher $\gamma_{max}’$ for the electron 
population, which
led to a slightly 
longer decay
time scale for corresponding 
pulses at
all energy bands (see Figure \ref{fig:LCs}). For the high-flux state of Mrk 421, the LC profiles are characterized by an even longer rise and decay time for pulses at all energies. In this case, the best fit SED was obtained using a higher value of $\Gamma^{'}_{sh}$ and a much harder spectral index compared to that of the low- and intermediate-flux states. Hence, the location of synchrotron and SSC peak frequencies shifted to higher values at 1 keV and 2 TeV, respectively. Since a relatively harder population of electrons is involved in producing the emission at 2 TeV compared to the other two 
days, the flare
at 2 TeV lasts for a longer time and decays very gradually, as shown in Figure \ref{fig:LCs}(d). Figure \ref{fig:hyst} shows the spectral evolution of the source with flux for various spectral states during the evolution of the flare. As can be seen, for all spectral states the source exhibits a clockwise pattern. This implies a soft lag and rapid variability at higher frequencies than at lower ones. 
It happens because higher energy electrons cool faster than the lower energy ones and the information about particle injection propagates from higher to lower energies through the electron population. This results in a characteristic clockwise loop because the injection is faster than the cooling. The loop depicts the manner in which the information on particle injection travels through the electron population at certain frequencies (\citealt{kirk1998}). Such clockwise patterns have previously been observed in Mrk 421 (\citealt{2009A&A...501..879T}; \citealt{1996ApJ...470L..89T}), and other sources, such as OJ 287 (\citealt{1986ApJ...304..295G}), and PKS 2155-304 (\citealt{1993ApJ...404..112S}). This also implies that
the evolution of the radiating particles is cooling dominated, as can be seen from the LCs in Figure \ref{fig:LCs} and has been previously reported by \citealt{singh2012NA}, for this time period. 
However, the spectral evolution at 10 keV 
for SS, low-flux, and intermediate-flux states
exhibits a brief counter-clockwise pattern before returning back to the clockwise pattern. This indicates a brief instant of acceleration and cooling timescales of particles becoming almost equal while producing 10 keV photons. As a result, there is a momentary decrease in the flux at 10 keV accompanied by spectral softening before the pulse at 10 keV returns to its original clockwise behavior.

\section{Conclusions:}

In this work, we carried out extensive simulations to study the spectral energy distributions of Mrk 421 during an outburst in mid-February 2010. During this period the source showed intense flux variations as it underwent a change from the low-flux state (2010 February 13-15) to an intermediate-flux state (observed on 2010 February 16) to a high-flux state (observed on 2010 February 17). In order to perform this study, we analyzed the optical/UV and GeV data and collected X-ray and TeV data from the literature for this period. 
Neecessary steps have been taken to correct for dereddening and host galaxy contribution for optical-UV data.
In addition, we compared the data for the low-flux state with a steady state of the source, which is an average of the emission over a period of 4.5 months. We used a time-dependent leptonic jet model in the internal shock scenario (MUZORF) to reproduce the above mentioned spectral states of Mrk 421. This allowed us to study the origin of variability on a daily basis and the role of intrinsic parameters in shaping the spectral states of the source. 
We note that the successful models are not unique reproductions of the observed data. Nevertheless, they provide a well-constrained parameter space for the values of key physical parameters, which was extracted using nearly simultaneous data from optical to VHE $\gamma$-rays using ground- and space-based facilities.
Our findings can be summarized as following:

(i) The key physical parameters that 
govern
the flaring activity in Mrk 421 for the time period under consideration are 
the kinetic luminosity
injected into the system that increases as the flare rises up, magnetic field strength that decreases during this evolution 
while the source departs progressively from equipartition,
energy index of the particle population that becomes harder, and the energy cutoffs of the particle population that increase with the rising of the flare. This type of a trend in the evolution of certain parameters could be taken as the general recipe of driving a multi-waveband flare in SSC-dominated sources. In addition, a change of particle acceleration 
mechanism, within the shock model
scenario, from Fermi first order to stochastic during the evolution of the flare might be required for driving such events in TeV blazars. 
However, we note that Fermi first-order mechanism under relativistic magnetic reconnection scenario has been suggested to produce flat electron energy spectra (\citealt{2014PhRvL.113o5005G}).

(ii) The low-flux state exhibited slightly different spectral features compared to SS and was identified with the onset of the flare. As the source evolved from a low- to a high-flux state it exhibited a "harder when brighter" behavior, which is common for Mrk 421.

(iii) A leptonic 
time-dependent synchrotron-SSC model
with multi-slice scheme successfully reproduces all the above-mentioned spectral states of Mrk 421. As also pointed out in \citealt{1995Malcomb}, the hadronic
model falls short of explaining the variability behavior (strong variability in X-ray \& TeV energy regimes but absence of that in the optical and MeV- few-Gev regimes) observed in Mrk 421 during this time period. On the other hand, MUZORF successfully explains this behavior, as also demonstrated in simulated LCs of the source (see Figure \ref{fig:LCs}), without invoking high-energy budget requirements associated with accelerating relativistic hadrons. 
However, it does not carry forward the information of one spectral state to the next and reproduces each of the states individually.

(iv) A simple power-law and an inhomogeneous emission region has been used to reproduce the spectral states during this flaring event. This is in contrast to some previous studies of this source, which mostly used a broken power-law and either a single homogeneous emission region or two disjointed emitting volumes in the jet (\citealt{HAGAR}; \citealt{TACTIC}; \citealt{SteadyState}; \citealt{2015Alexic}; \citealt{ARGO}; \citealt{doi:10.1093/mnras/stx2185}). This could imply that inhomogeneous emission regions give a better agreement with the observed data without invoking complicated distributions of particle population. However, we note that a particle population following a log-parabolic 
distribution gives
an even better agreement with the broadband SED of Mrk 421 for such flaring events. This is because such distributions are a natural consequence of stochastic particle acceleration and reproduce the behavior of 
particle populations
more accurately than a simple power-law distribution (\citealt{massaro2006}; \citealt{1962SvA.....6..317K}; \citealt{IC}). 

(v) The successful model for the high-flux state is characterized 
by a harder
electron energy index with electron population distributed at higher frequencies, higher Doppler factor, and low magnetic field values. A low magnetic field takes care of the required high-flux level of the SSC component observed that day and at the same time suppresses the synchrotron component. The successful model indicates that such "harder when brighter" spectral states are successfully reproduced by a set of parameters instead of a single physical parameter.

(vi) As shown in Figure \ref{fig:MWL}, the observed data exhibits no variability in the optical, mild variability in the Fermi-GeV regime, and significant variability at X-rays and TeV energies for this time period. This is supported by the LC realizations of our successful models of different spectral states (see Figure \ref{fig:LCs}). The LC profiles of both optical and GeV energies peak later compared to X-ray and TeV flares for all days because 
relatively low-energy
electrons are involved in their production via synchrotron and SSC processes, respectively. Hence, these flares peak later and decay gradually compared to their X-ray and TeV counterparts.

(vii) The flare at 2 TeV is produced via 
the SSC mechanism.
As a result, it peaks later compared to its synchrotron-dominated X-ray flares and decays gradually. However, for the high-flux state the flare 
exhibits
a further spectral hardening and a shifting of the peak of the SSC component to the 2 TeV energy regime. As a result, the flare 
lasts
for longer and 
decays
much more gradually compared to 2 TeV flares at other days.

(viii) Due to the fact that the flaring episode could not be observed with 
high temporal
resolution, the simulated LCs can not be validated with the observed LCs. A 
night-long simultaneous data set (like 
the campaign
described in \citealt{MAGICNuStar}) with variability signatures might provide further scope to validate the model predictions.

\section{Acknowledgements}
In this research data, software and web tools obtained from High Energy
Astrophysics Science Archive Research Center (HEASARC) have been used. This is a service provided by Goddard Space Flight Center and the Smithsonian Astrophysical Observatory. We acknowledge the computational resources provided to us by Boston University through the Massachusetts Green High Performance Computing Center. 
M.J. acknowledges support by NASA through Fermi grant 80NSSC17K0650.
Data from the Steward Observatory spectropolarimetric monitoring project were used. 
Support for the analysis of optical data was provided by National Science Foundation grant AST-1615796. 
We thank ground- and space-based monitoring facilities used in this work for their data without which this study would not have been possible.

\bibliographystyle{mnras} 
\bibliography{SourceFiles/ref}
\end{document}